\documentclass[pra,aps,10pt,superscriptaddress,twocolumn,floatfix]{revtex4-1}  
\usepackage{graphicx,color}
\usepackage{amsmath,amssymb,bm}
\usepackage{braket}		

\DeclareMathOperator{\Tr}{Tr}

\usepackage[plainpages=false,pdfpagelabels,colorlinks=true,linkcolor=red,urlcolor=blue,citecolor=blue,pdftitle={Titl}e,pdfauthor={},pdfdisplaydoctitle=true,pdfduplex=DuplexFlipLongEdge]{hyperref}

\definecolor{darkred}{rgb}{0.90,0.2,0.2}
\definecolor{darkgreen}{rgb}{0,0.60,.2}
\definecolor{darkblue}{rgb}{0.1,0.3,1}
\definecolor{grey}{cmyk}{0,0,0,0.25}
\definecolor{orange}{cmyk}{0,0.6,0.8,0}

\begin{document}
\title{Impenetrable SU($N$) fermions in one-dimensional lattices}

\author{Yicheng Zhang}
\affiliation{Department of Physics, The Pennsylvania State University, University Park, Pennsylvania 16802, USA}
\affiliation{Department of Theoretical Physics, J. Stefan Institute, Ljubljana, Slovenia}
\author{Lev Vidmar}
\affiliation{Department of Theoretical Physics, J. Stefan Institute, Ljubljana, Slovenia}
\author{Marcos Rigol}
\affiliation{Department of Physics, The Pennsylvania State University, University Park, Pennsylvania 16802, USA}

\begin{abstract}
We study SU($N$) fermions in the limit of infinite on-site repulsion between all species. We focus on states in which every pair of consecutive fermions carries a different spin flavor. Since the particle order cannot be changed (because of the infinite on-site repulsion) and contiguous fermions have a different spin flavor, we refer to the corresponding constrained model as the model of distinguishable quantum particles. We introduce an exact numerical method to calculate equilibrium one-body correlations of distinguishable quantum particles based on a mapping onto noninteracting spinless fermions. In contrast to most many-body systems in one dimension, which usually exhibit either power-law or exponential decay of off-diagonal one-body correlations with distance, distinguishable quantum particles exhibit a Gaussian decay of one-body correlations in the ground state, while finite-temperature correlations are well described by stretched exponential decay.
\end{abstract}

\maketitle


\section{Introduction}

Low dimensionality can give rise to fascinating phenomena. A striking one, which occurs when interacting spinful particles are confined in one dimension, is the so-called spin-charge separation~\cite{tomonaga_50, luttinger_63}. This phenomenon has been studied experimentally in a wide range of systems including quasi-one-dimensional compounds~\cite{kim_matsuura_96, segovia_purdie_99, claessen_sing_02, kim_koh_06}, grain boundaries in semiconductors~\cite{ma_diaz_17}, and ultracold atoms in optical lattices~\cite{boll_hilker_16, hilker_17}. Theoretically, a cornerstone in the understanding of the low-energy properties of one-dimensional (1D) systems is provided by the Luttinger liquid theory~\cite{haldane_81, voit_95, giamarchibook, cazalilla_citro_review_11}, which, among others, naturally describes spin-charge separation and predicts the existence of power-law correlations in gapless ground states.

Recent impetus for the study of many-body quantum systems in one dimension has been provided by experimental advances with ultracold quantum gases in optical lattices~\cite{bloch08, cazalilla_citro_review_11}. In optical lattices, it is possible to realize very strong correlations in one dimension so that, e.g., bosons with contact interactions behave as impenetrable particles and ``fermionize''~\cite{girardeau_60, kinoshita_wenger_04, paredes_widera_04, kinoshita_wenger_05}. Another remarkable possibility presented by experiments with ultracold gases is the use of fermionic alkaline earth atoms to realize SU$(N)$ models~\cite{Wu_2003, cazalilla_2009, Gorshkov_2010, Taie_2012, Pagano_2014, Scazza_2014, Zhang_2014} (see Ref.~\cite{cazalilla_rey_14} for a review). The latter possibility has motivated theoretical studies on spin chains and quantum gases beyond the more traditionally considered SU(2) case~\cite{nataf_mila_14, volosniev_14, dufour_nataf_15, decamp_armagnat_16, decamp_junemann_16, laird_shi_17, jen_yip_18}. 

Both the realization of SU$(N)$ models and the achievement of regimes in one dimension in which particles become ``impenetrable''  are important motivations for this work. In the limit of infinite contact repulsion between spinful particles, the spin and charge degrees of freedom decouple at all energies. This can be used to construct many-body wavefunctions~\cite{ogata_shiba_90}. Physical properties of  spinful impenetrable particles have been studied within different models, ranging from fermionic models~\cite{parola_sorella_90, parola_sorella_92, penc_96, penc_hallberg_97, izergin_pronko_98, prelovsek_elshawish_04, kumar_08, kumar_09, bertini_tartaglia_17, essler-book, jen_yip_18} to classical systems~\cite{medenjak_klobas_17}.

\begin{figure}[!t]
\includegraphics[width=0.99\columnwidth]{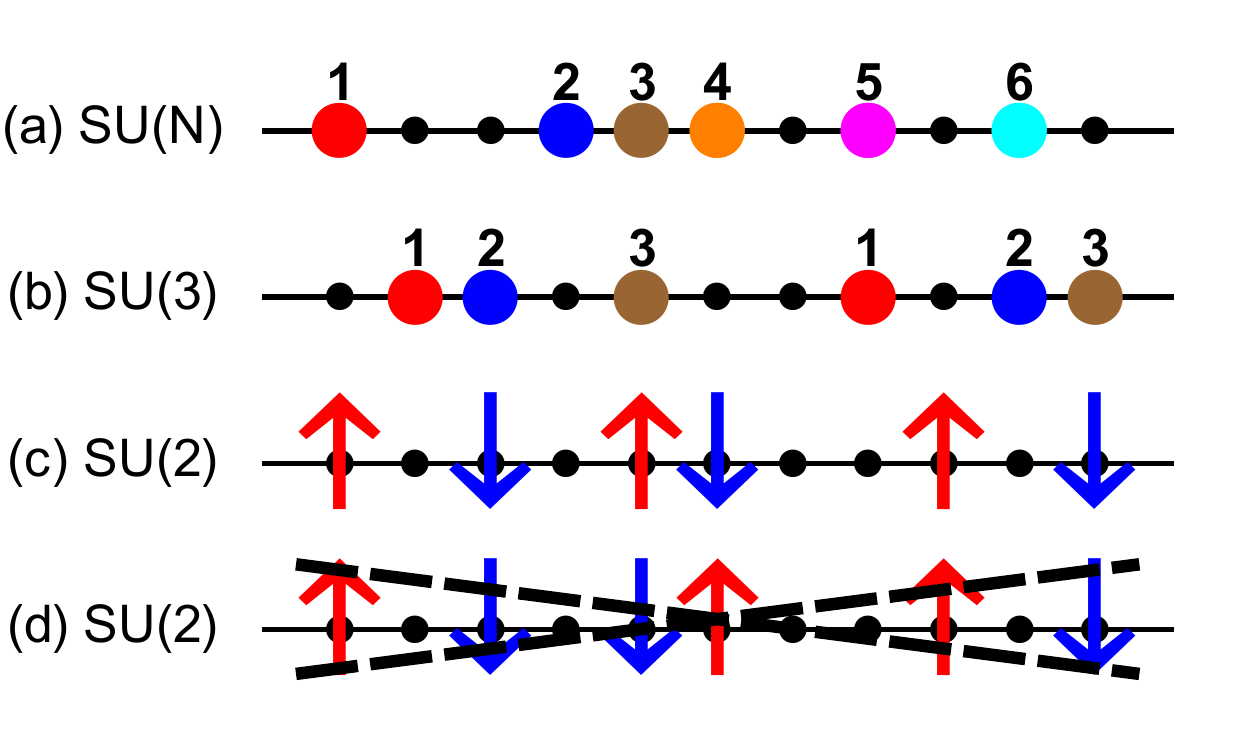}
\caption{Particle configurations of SU($N$) fermions on a chain with $N_p = 6$ particles. (a)--(c) Configurations in which every pair of consecutive particles carries a distinct spin flavor, and in which there is an identical spin flavor pattern that we call generalized Neel order. Configurations (a)--(c) can be described by the model of distinguishable quantum particles introduced in this work. (d) A configuration that cannot be described by the model of distinguishable quantum particles.} 
\label{fig1}
\end{figure}

Here we study impenetrable SU($N$) fermions, which can be thought of as the limit of infinite on-site repulsion between all spin flavors in a generalized 1D SU($N$) Hubbard model. In this limit, sectors of the Hamiltonian with different spin configurations are degenerate. Such independent sectors are in principle accessible in experiments with ultracold atoms in optical lattices via spin-resolved manipulation techniques to imprint desirable spin patterns~\cite{hild_fukuhara_14}. In this study, we focus on sectors in which contiguous fermions have different spin flavors. For eigenstates of the Hamiltonian in those sectors, we introduce an exact numerical method to calculate one-body correlations via a mapping onto noninteracting spinless fermions (SFs). This method is used to study properties of  ground states with special spin flavor patterns, namely, with generalized Neel order. Figures~\ref{fig1}(a)--\ref{fig1}(c) display examples of many-particle configurations of interest in this work, while Fig.~\ref{fig1}(d) displays an example which does not satisfy the requirement of contiguous fermions having different spin flavors. 

Since particle exchange is forbidden in the limit of infinite on-site repulsion between all spin flavors, and contiguous fermions have a different spin flavor in the many-particle configurations of interest here, we refer to the corresponding constrained model as a model of {\it distinguishable} quantum particles (DQPs). Note that the ground state and finite-temperature states (also studied in this work) of DQPs are not the ground state and finite-temperature states of the SU($N$) Hubbard model in the limit of infinite on-site repulsion. The latter involve states with exponentially many spin configurations. SU($N$) fermions with infinite on-site repulsion become distinguishable only if one constrains the system to be in a sector with a specific spin configuration in which contiguous fermions have a different spin flavor.

Another motivation for this study is the fact that our constrained SU($N$) model exhibits fundamentally different off-diagonal one-body correlations compared to unconstrained models. For the latter, the exact Bethe ansatz solution for the 1D SU(2) Hubbard model was obtained by Lieb and Wu~\cite{lieb_wu_68}, and simplified by Ogata and Shiba in the limit of infinite repulsion~\cite{ogata_shiba_90}. Here, we show that one-body correlations in the ground state of DQPs exhibit a Gaussian decay with distance, in contrast to the power-law decay of the unconstrained SU(2) case~\cite{frahm_korepin_90, penc_hallberg_97}. At finite temperatures, one-body correlations are shown to be well described by a stretched exponential decay, with an exponent that transitions (with increasing temperature) between the Gaussian decay at zero temperature and an exponential decay at high temperatures.

The presentation is organized as follows. In Sec.~\ref{sec2}, we introduce the constrained SU($N$) model for distinguishable quantum particles and the methodology developed to evaluate its one-body correlations. Numerical results for these correlations are presented in Sec.~\ref{sec3} for the ground state and in Sec.~\ref{sec4} for finite temperatures. A summary of the results is presented in Sec.~\ref{sec5}.

\section{Setup and Formalism} \label{sec2}

We start with a generalized 1D Hubbard model for SU($N$) fermions with infinite on-site repulsion. The model Hamiltonian for a chain with open boundaries can be written as
\begin{equation}
\label{def_H}
 \hat H_N = -J\sum_{l=1}^{L-1}\sum_{\sigma=1}^{N}\left[\hat f^{(\sigma)\dagger}_l \hat f^{(\sigma)}_{l+1}+\hat f^{(\sigma)\dagger}_{l+1} \hat f^{(\sigma)}_l\right]\,,
\end{equation}
where $L$ is the number of lattice sites, $\sigma$ denotes the spin flavor (in our notation, $\sigma \in \{ 1, ..., N \}$), and $\hat f^{(\sigma)\dagger}_l$ ($\hat f^{(\sigma)}_l$) is the creation (annihilation) operator for a fermion with spin $\sigma$ at site $l$. Infinite on-site repulsion is enforced by the constraints $\hat f^{(\sigma)\dagger}_l \hat f^{\dagger(\sigma')}_l = \hat f^{(\sigma)}_l \hat f^{(\sigma')}_l=0$. The hopping amplitude $J$ and the lattice spacing are set to unity.

The Hamiltonian $\hat H_N$ in Eq.~(\ref{def_H}) commutes with the total particle number operator for any given spin flavor $\hat N^{(\sigma)}_p = \sum_{l=1}^{L} \hat f^{(\sigma)\dagger}_l \hat f^{(\sigma)}_{l}$ (the total number of particles with any given spin flavor is conserved) and, hence, with the total particle number operator $\hat N_p = \sum_{\sigma=1}^{N} \hat N^{(\sigma)}_p$. Moreover, as a consequence of the infinite on-site repulsion, $\hat H_N$ also preserves any configuration of spin flavors. Hence, for a given total particle number $N_p$, the Hamiltonian consists of degenerate sectors, where every block is associated with the spin configuration $\underline{\sigma} = \{ \sigma_1, ..., \sigma_{N_p} \}$ ($\sigma_j \in \{ 1, ..., N \}$). Eigenstates within a block are linear superpositions of base kets of the form
\begin{equation}
\label{neel_state}
|\varphi_{\underline{x},\underline{\sigma}} \rangle = \prod_{j=1}^{N_p} \hat f_{x_j}^{(\sigma_j)\dagger} | \emptyset \rangle \, ,
\end{equation}
where $\underline{x} = \{ x_1, ..., x_{N_p} \}$ denotes the set of occupied sites, $x_j \in \{1, ..., L \} $, and $x_1 < x_2 < ... < x_{N_p}$.

We study the impenetrable SU($N$) model, with Hamiltonian $\hat H_N$ in Eq.~(\ref{def_H}), within a sector with a given spin configuration $\underline{\sigma}$. This is a model of relevance to experiments with ultracold quantum gases in optical lattices in which such spin configurations can be constructed using, e.g., spin-resolved manipulation techniques (which can be applied to simple product states~\cite{hild_fukuhara_14}) followed by adiabatic or quasiadiabatic transformations.

\subsection{Model of distinguishable quantum particles}

The next essential constraint imposed on the states we study, in addition to being for impenetrable SU($N$) fermions within a single spin configuration $\underline{\sigma}$ sector, is that we require the spin configuration $\underline{\sigma}$ to have every pair of consecutive fermions carry distinct spin flavors:
\begin{equation} \label{def_dqp}
 \underline{\sigma} = \{ \{ \sigma_j \} \,;\, j=1,...,N_p \, \,;\, \sigma_j \neq \sigma_{j+1} \, \forall \, j < N_p \} \, .
\end{equation}
This implies that particle exchanges are forbidden. We call the constrained SU($N$) model, in which the spin configuration obeys the condition in Eq.~(\ref{def_dqp}), a model of distinguishable quantum particles.

Of particular interest to us is the DQP model in which the spin configuration forms a periodic structure, which we call generalized Neel order. These configurations have
\begin{equation} \label{def_neel}
 \underline{\sigma} = \{ \{ \sigma_j \} \,;\, j=1,...,N_p \, \,;\, \sigma_j = \left[ (j-1) \, \mbox{mod} \, N \right]  + 1\} \ , 
\end{equation}
Examples of such configurations are schematically shown in Fig.~\ref{fig1}. Note that, in the SU(2) case, the spin configurations that obey Eq.~(\ref{def_dqp}) also obey Eq.~(\ref{def_neel}).

At this point it is important to stress that while we have arrived at the model of DQPs thinking about experiments with impenetrable SU($N$) fermions, the approach we develop in what follows and the results we obtain apply equally to spinful impenetrable bosons. Under the constraints of our construction, the original particle statistics plays no role.

\subsection{Spin-charge decoupling}

For a given spin configuration $\underline{\sigma}$, charge degrees of freedom of the constrained SU($N$) model can be described by the spinless fermion Hamiltonian
\begin{equation}
\label{def_Hsf}
 \hat H_{\rm SF} = -\sum_{l=1}^{L-1}{(\hat c^{\dagger}_l \hat c^{}_{l+1}+\hat c^{\dagger}_{l+1} \hat c^{}_l)}\,,
\end{equation}
where $\hat c^\dagger_l$ ($\hat c_l$) is the spinless fermion creation (annihilation) operator at  lattice site $l$. The challenge that remains is to take into account the spin degrees of freedom to compute spin-resolved off-diagonal correlation functions.

Our solution to this challenge is based on the following ansatz within the model of DQPs for the spin-resolved one-body correlations $\hat C^\sigma_l(x) = \hat f^{(\sigma)\dagger}_{l+x} \hat f^{(\sigma)}_l$, between site $l$ and site $l+x$,
\begin{equation}
\label{corr_map}
C^\sigma_l(x)=\langle\Psi| \hat f^{(\sigma)\dagger}_{l+x} \hat f^{(\sigma)}_l |\Psi\rangle=\langle\Psi_{\rm SF}| \hat c^{\dagger}_{l+x} \hat c^{}_l\, {\cal \hat P}^{(\sigma)}_{l,x} |\Psi_{\rm SF}\rangle \,,
\end{equation}
where $|\Psi\rangle$ and $|\Psi_{\rm SF}\rangle$ are eigenstates (we focus on the ground state later) of $\hat H_N$ (in the sector with the desired spin order) and $\hat H_{\rm SF}$, respectively. ${\cal \hat P}^{(\sigma)}_{l,x}$ is a spin projection operator that, acting on $|\Psi_{\rm SF}\rangle$, produces a (polynomially large in the system size) sum of Slater determinants. The final expression in Eq.~\eqref{corr_map} can be efficiently evaluated in polynomial time using properties of Slater determinants, as done for hard-core bosons in Refs.~\cite{rigol_muramatsu_04sept} and \cite{rigol_muramatsu_05july}.

\subsection{Spin projection operator} \label{sec2_projector}

We construct the projection operator ${\cal \hat P}^{(\sigma)}_{l,x}$ as the product of two operators:
\begin{equation} \label{def_projP}
{\cal \hat P}^{(\sigma)}_{l,x} = {\cal \hat M}_{l,x}{\cal \hat R}^{(\sigma)}_{l}\,.
\end{equation}
The role of the operator ${\cal \hat M}_{l,x}$ is to prevent an exchange of particles that would result in a change of the spin configuration in the SU($N$) model. This operator must annihilate many-body states in which any of the lattice sites $j \in \{l+1,...,l+x-1 \}$ is occupied. This is achieved by defining
\begin{equation} \label{def_projM}
 {\cal \hat M}_{l,x}=\prod_{j=l+1}^{l+x-1} \left(1- \hat c_j^\dagger \hat c^{}_j \right) \, .
\end{equation}

On the other hand, the role of the operator ${\cal \hat R}^{(\sigma)}_{l}$ is to target the spin flavor $\sigma$ at site $l$. Let us first focus on spin configurations with generalized Neel order, Eq.~(\ref{def_neel}). (We consider arbitrary spin configurations within the DQP model right afterward.) We define ${\cal \hat R}^{(\sigma)}_{l}(N)$ as
\begin{equation} \label{def_projR}
{\cal \hat R}^{(\sigma)}_{l}(N) = \frac{1}{N} \sum_{k=0}^{N-1}{e^{-\frac{2\pi i}{N}\sigma k}\exp\left[\frac{2\pi i}{N}k\sum_{j=1}^l \hat c_j^\dagger \hat c^{}_j \right]} \, .
\end{equation}
This operator, which involves counting the particles at sites $1$ through $l$, ensures that the number of particles between site $1$ and site $l$ is the appropriate one for the given spin flavor $\sigma$ to occur at site $l$. To prove it, one can express the wave function $|\Psi_{\rm SF}\rangle$ as a sum of many-body states $|\phi^l_a \rangle$, where each $|\phi^l_a \rangle$ is a linear superposition of base kets that share a common property, namely, that the total number of particles at sites $1$ through $l$ is $a$. 

We focus on the case in which $L-l\geq N_p-1$ (similar formulas can be derived for $L-l<N_p-1$). The decomposition in terms of $|\phi^l_a \rangle$ implies that
\begin{equation} \label{project_number}
\sum_{j=1}^l \hat c_j^\dagger \hat c^{}_j|\Psi_{\rm SF}\rangle=\sum_{m=1}^{m_{\rm max}} \sum_{\sigma'=1}^{\sigma_{\rm max}(m)}[(m-1)N+\sigma']|\phi^l_{(m-1)N+\sigma'}\rangle,
\end{equation} 
where $m_{\rm max}$ and $\sigma_{\rm max}(m)$ are such that all the possible particle numbers at sites $1$ through $l$ are included. If $l\leq N_p$, then $m_{\rm max} =\lceil l/N\rceil$, $\sigma_{\rm max}(m<m_{\rm max}) = N$, and $\sigma_{\rm max}(m_{\rm max}) = l -(m_{\rm max}-1)N$. On the other hand, if $l>N_p$, then $\sigma_{\rm max}(m) = N$ and $m_{\rm max} = N_p/N$. We assume that $N_p/N$ is an integer. 

Using Eq.~(\ref{project_number}), the projector ${\cal \hat R}^{(\sigma)}_{l}(N)$ acting on $| \Psi_{\rm SF}\rangle$ yields
\begin{align}
\hspace*{-0.1cm}\label{def_R_psi}
 & {\cal \hat R}^{(\sigma)}_{l}(N) | \Psi_{\rm SF}\rangle \\
 & = \sum_{m=1}^{m_{\rm max}} \sum_{\sigma'=1}^{\sigma_{\rm max}(m)} \frac{1}{N} \sum_{k=0}^{N-1} e^{-\frac{2\pi i}{N}(\sigma-\sigma') k } |\phi^l_{(m-1)N+\sigma'}\rangle \nonumber \\
 & = \sum_{m=1}^{m_{\rm max}} \sum_{\sigma'=1}^{\sigma_{\rm max}(m)} \delta_{\sigma,\sigma'} |\phi^l_{(m-1)N+\sigma'}\rangle \, = \, \sum_{m=1}^{m'_{\rm max}} |\phi^l_{(m-1)N+\sigma}\rangle \, ,
 \nonumber 
\end{align} 
where $m'_{\mathrm{max}}=m_{\mathrm{max}}$ if $\sigma_{\mathrm{max}}(m_{\mathrm{max}})\geq \sigma$, and $m'_{\mathrm{max}}=m_{\mathrm{max}}-1$ otherwise. We then see that ${\cal \hat R}^{(\sigma)}_{l}(N) | \Psi_{\rm SF}\rangle$ results in states with the desired numbers of particles between site 1 and site $l$, so that if there is a particle at site $l$ it must have flavor $\sigma$.

Equation~(\ref{def_projR}) can be rewritten in a simple form for the SU($2$) case, where the spin is either up ($\sigma=1$) or down ($\sigma=2$). In this case
\begin{equation}
{\cal \hat R}^{(\sigma)}_{l}(2) = \frac{1}{2} \left(1 + (-1)^{\sigma} \exp\left[i \pi \sum_{j=1}^l \hat c_j^\dagger \hat c^{}_j \right]\right) \,.
\end{equation}

Another interesting limit is the case in which $N=N_p$ so that every particle carries a distinct spin flavor. As a consequence, every $\sigma$ can be uniquely assigned to a particle $j$, resulting in $\sigma_j$. Then, the projector ${\cal \hat R}^{(\sigma)}_{l}(N_p) = {\cal \hat R}^{(\sigma_j)}_{l}(N_p)$ can target any particle $j \in \{ 1, ..., N_p \}$. We denote one-body correlations in that case as $C_l^{\sigma_j}(x)$. These correlations can be used to compute $C^\sigma_l(x)$ for any eigenstate of $\hat H_N$ in a sector with a desired spin order obeying the condition in Eq.~(\ref{def_dqp}):
\begin{equation} \label{c_dqp}
C_l^\sigma(x) = \sum_{\sigma_j = \sigma} C_l^{\sigma_j}(x) \, .
\end{equation}
Note that the previous expression can also be used for spin configurations exhibiting the generalized Neel order in Eq.~\eqref{def_neel}. However, given the operators defined in Eqs.~\eqref{def_projP}--\eqref{def_projR}, it would be inefficient computationally to use Eq.~\eqref{c_dqp} for states exhibiting such an order.

\subsection{Universality of the total one-body correlations}

For a system with an arbitrary number of flavors $N$, and an arbitrary configuration of the spins obeying the condition in Eq.~(\ref{def_dqp}), it is of interest to determine the total one-body correlation function (the sum over all spin flavors)
\begin{equation} \label{c_total}
 C_l(x)= \sum_{\sigma=1}^{N} C_l^{\sigma}(x) \, .
\end{equation}
This can be done by combining Eqs.~(\ref{c_dqp}) and~(\ref{c_total}), so that the sum over distinct spin flavors is replaced by the sum over all particles, $C_l(x) = \sum_{\sigma_j = 1}^{N_p} C_l^{\sigma_j}(x)$. It yields
\begin{align}
\label{avg_corr}
 C_l(x) =& \nonumber \frac{1}{N_p}\sum_{k=0}^{N_p-1} \sum_{\sigma_j=1}^{N_p} e^{-\frac{2\pi i}{N_p}\sigma_j k} \\
   & \times  \langle \hat c^{\dagger}_{l+x} \hat c_l\, {\cal \hat M}_{l,x} \exp\left[\frac{2\pi i}{N_p}k\sum_{j=1}^l \hat c_j^\dagger \hat c^{}_j\right]\rangle \nonumber \\
 = &\sum_{k=0}^{N_p-1} \delta_{k,0} \langle \hat c^{\dagger}_{l+x} \hat c_l\, {\cal \hat M}_{l,x} \exp\left[\frac{2\pi i}{N_p}k\sum_{j=1}^l \hat c_j^\dagger \hat c^{}_j\right]\rangle \nonumber \\
 = & \langle \hat c^{\dagger}_{l+x} \hat c_l\, {\cal \hat M}_{l,x} \rangle \, .
\end{align}

Equation~(\ref{avg_corr}) shows that the total one-body correlations are independent of the number of flavors and of the particular spin pattern selected, as long as the condition in Eq.~(\ref{def_dqp}) is satisfied. Only the projector ${\cal \hat M}_{l,x}$ is needed when computing $C_l(x)$. The result in Eq.~(\ref{avg_corr}) is one of our motivations for calling the impenetrable SU($N$) model under the constraints imposed on the spin configurations a model for DQPs.

\section{Ground state} \label{sec3}

We now turn our attention to the DQP model with spin configurations that exhibit generalized Neel order, Eq.~(\ref{def_neel}). Here we study ground-state properties. In Sec.~\ref{sec4}, we study finite-temperature properties.

\subsection{Numerical implementation}

We use a numerical procedure based on properties of Slater determinants to calculate one-body correlation functions (analogous to the one introduced in Refs.~\cite{rigol_muramatsu_04sept} and \cite{rigol_muramatsu_05july} for hard-core bosons). We express Eq.~(\ref{corr_map}) as
\begin{equation} \label{eq:opdm1}
C^\sigma_l(x)=\delta_{x,0} \langle\Psi^G_{\rm SF}|{\cal \hat P}^{(\sigma)}_{l,x}|\Psi^G_{\rm SF}\rangle-\langle\Psi^G_{\rm SF}|  \hat c^{}_{l} \hat c^{\dagger}_{l+x} {\cal \hat P}^{(\sigma)}_{l,x} |\Psi^G_{\rm SF}\rangle \,,
\end{equation}
where the ground state for spinless fermions is a Slater determinant, which can be written as $|\Psi^G_{\rm SF}\rangle = \prod_{j=1}^{N_p}\sum_{m=1}^L G_{mj} \hat c^\dagger_m|\emptyset\rangle$. The projection operator defined in Eqs.~(\ref{def_projP})-(\ref{def_projR}) changes the spinless fermion ground state into the linear combination of Slater determinants
\begin{equation}
{\cal \hat P}^{(\sigma)}_{l,x}|\Psi^G_{\rm SF}\rangle=\frac{1}{N}\sum_{k=0}^{N-1} e^{-\frac{2\pi i}{N}\sigma k}\prod_{j=1}^{N_p}\sum_{m=1}^L G^k_{mj} \hat c^\dagger_m|\emptyset\rangle \,,
\end{equation}
with
\begin{equation} \label{def_Gijk}
G^k_{mj}= \Bigg \{
	\begin{tabular} {ccl}
	$e^{\frac{2\pi i}{N}k}\,G_{mj}$,&&$m\leq l$;\\
	$0$, && $l<m<l+x$;\\
        $G_{mj}$ && ${\rm otherwise.}$
	\end{tabular} \, 
\end{equation}
The modifications in $G^k_{mj}$ with respect to $G_{mj}$ are due to ${\cal \hat R}^{(\sigma)}_{l}(N)$ for $m\leq l$, and due to ${\cal \hat M}_{l,x}$ for $l<m<l+x$. Next, we use that $\hat c^\dagger_j$ acting on a Slater determinant, specified by a matrix ${\bf G}$ with $N_p$ columns, results in a new Slater determinant specified by a matrix ${\bf G}'$, which is just ${\bf G}$ with an added column, $G'_{m,N_p+1} = \delta_{m,j}$. This means that, to compute the second expectation value in Eq.~\eqref{eq:opdm1}, we need to change $\bf{G} \to {\bf G'}$ when acting with $\hat c^{}_{l}$ on the left, and ${\bf G}^k \to {\bf G}'^k$ when acting with $\hat c^{\dagger}_{l+x}$ on the right. (Here, $\bf G$ and ${\bf G}^k$ are matrices with elements $G_{ij}$ and $G_{ij}^k$, respectively.) The final step to evaluate Eq.~\eqref{eq:opdm1} is to compute the inner product of Slater determinants, which is equal to the determinant of the product of the matrices specifying the Slater determinants \cite{rigol_muramatsu_04sept, rigol_muramatsu_05july}. 

Putting all the above together, the one-body correlation function can be calculated as
\begin{equation}
C^\sigma_l(x)=\frac{1}{N}\sum_{k=0}^{N-1}e^{-\frac{2\pi i}{N}\sigma k}(\delta_{x,0}\det[{\bf G}^\dagger {\bf G}^k]-\det[{\bf G}'^\dagger {\bf G}'^k]) \,.
\end{equation}

The total one-body correlation function $C_l(x)$ is much simpler to calculate, following the procedure outlined above we get:
\begin{equation}
C_l(x)=\delta_{x,0}\det[{\bf G}^\dagger {\bf G}^{k=0}]-\det[{\bf G}'^\dagger {\bf G}'^{k=0}] \,.
\end{equation}

\subsection{Results for finite systems}

\begin{figure}[!t]
\includegraphics[width=0.99\columnwidth]{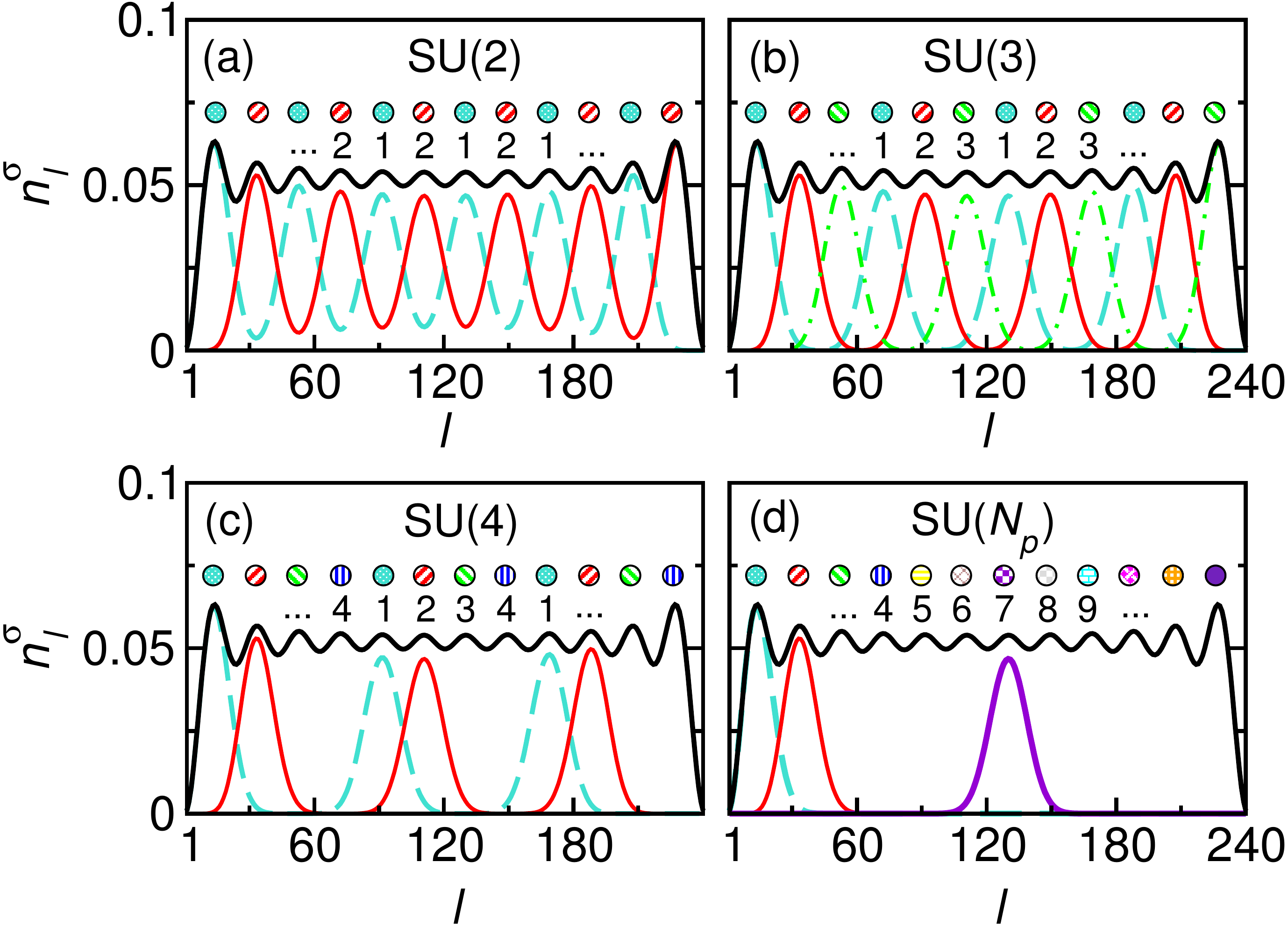}
\caption{Site occupations $n^\sigma_l$ in the ground state of the (a) SU(2), (b) SU(3), (c) SU(4), and (d) SU($N_p$) models, for $N_p=12$ particles in an open chain with $L=240$ sites. Circles highlight the spin flavors, with the numbers below them indicating the corresponding value of $\sigma$. In (c) and (d) we show $n^\sigma_l$ only for a few spin flavors. The thick solid line above the $n^\sigma_l$ profiles shows the total (the sum over all spin flavors) site occupations, which are identical to those of the corresponding spinless fermion Hamiltonian, Eq.~(\ref{def_Hsf}).} 
\label{fig2}
\end{figure}

In Fig.~\ref{fig2}, we plot the site occupations $n^\sigma_l \equiv C^\sigma_l(0)$ of different spin flavors in the ground state of $N_p = 12$ impenetrable fermions with a generalized Neel pattern on a lattice with $L=240$ sites. We show results for the SU(2), SU(3), SU(4), and SU($N_p$) cases in Figs.~\ref{fig2}(a), \ref{fig2}(b), \ref{fig2}(c), and \ref{fig2}(d), respectively. Note that, in these finite systems at low filling, particles have relatively well-defined regions of the lattice on which they can be found. Figure~\ref{fig2} also shows the total (the sum over all spin flavors) site occupations $n_l = \sum_\sigma n^\sigma_l$ (solid black lines above the spin-resolved site occupations). They are identical to the site occupations in the model of spinless fermions [Eq.~(\ref{def_Hsf})] onto which each constrained SU($N$) model is mapped. In finite systems, small peaks in $n_l$ are the remnants of the DQP positions. What happens in the thermodynamic limit for systems in which $N$ is $O(1)$, namely, when $N$ does not scale with $N_p$, is discussed in Sec.~\ref{sec2_2}.

\begin{figure}[!t]
\includegraphics[width=0.99\columnwidth]{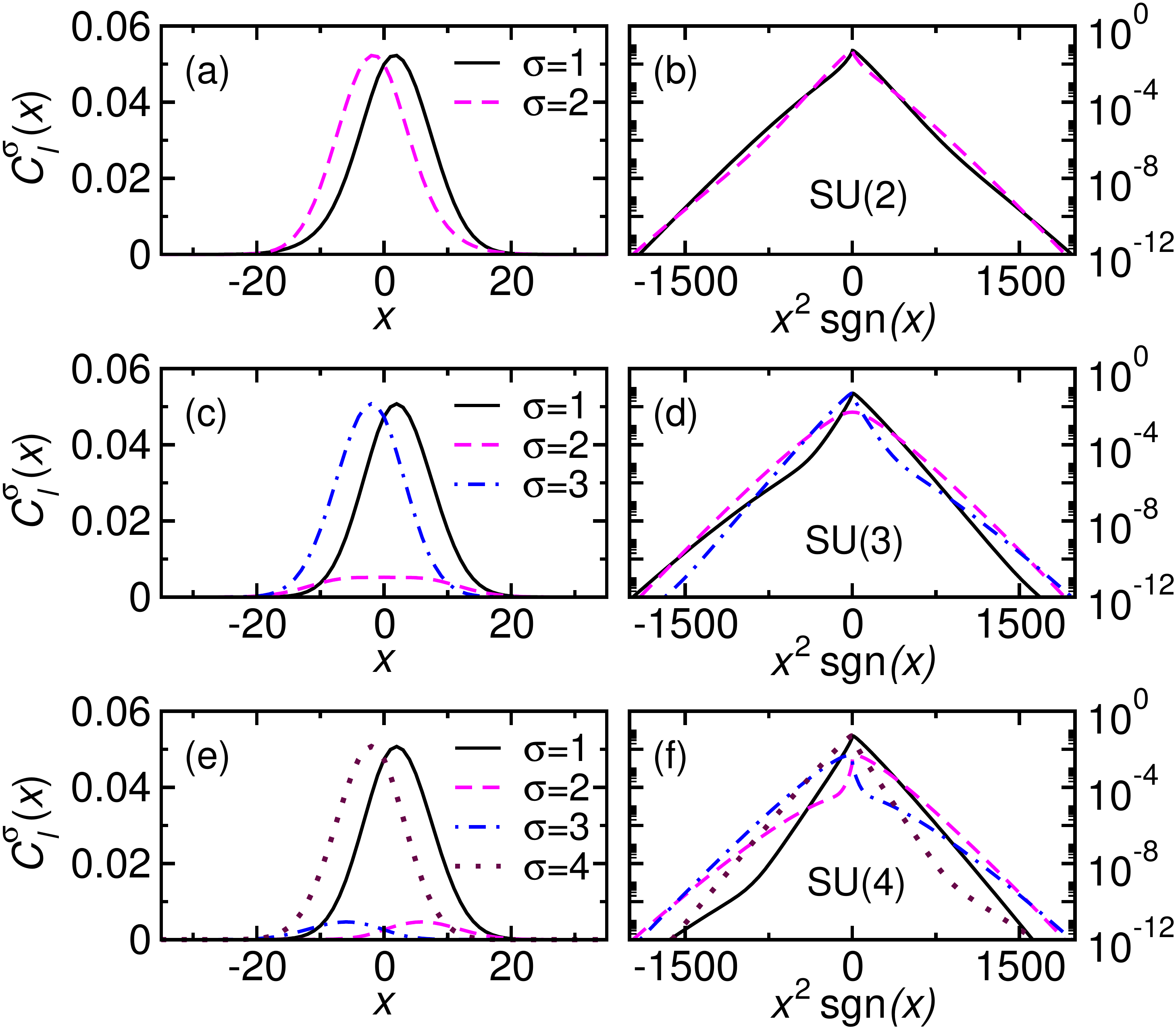}
\caption{One-body correlations $C^\sigma_l(x)$ in the ground state of the (a, b) SU(2), (c, d) SU(3), and (e, f) SU(4) models. The results are for open chains with $L=2401$ sites, the total number of particles $N_p=240$, and $l=1201$ (the site at the center of the chain). Results in (b), (d), and (f) are the same as those in (a), (c), and (e), respectively. The only difference is that the axes are rescaled.} 
\label{fig3}
\end{figure}

Fig.~\ref{fig3}(a), \ref{fig3}(c), and \ref{fig3}(e) show the behavior (on a linear scale) of the off-diagonal matrix elements of the one-body correlation matrix $C^\sigma_l(x)$, with $l$ being the site at the center of an open chain with $L=2401$. Results for $C^\sigma_l(x)$ are shown for all spin flavors in the SU(2), SU(3), and SU(4) models. $C^\sigma_l(x)$ can be seen to depend on $\sigma$, which is consistent with the observation in Fig.~\ref{fig2} that, in finite systems, particles (and hence flavors) can be found in relatively well-defined regions of the lattice.

Fig.~\ref{fig3}(b), \ref{fig3}(d), and \ref{fig3}(f) show the same $C^\sigma_l(x)$ as in the left panels but plotted on a log scale versus $x^2 {\rm sgn}(x)$. Most of the curves exhibit a near-linear decay with $x^2$, which indicates a near-Gaussian decay of one-body correlations. In Sec.~\ref{sec2_3}, we show that the total one-body correlations $C_l(x)$ exhibit a Gaussian decay in finite systems (even smaller than the ones in Fig.~\ref{fig3}). In Sec.~\ref{sec2_2}, we argue that $C^\sigma_l(x)$ exhibits a Gaussian decay in the thermodynamic limit in systems in which $N$ is $O(1)$.

An observable that is of special interest for experiments with ultracold gases in optical lattices is the quasimomentum distribution function
\begin{equation}\label{eq:nk}
m^\sigma_k=\frac{1}{L}\sum_{l,x}e^{-ikx}C^\sigma_l(x) \, ,
\end{equation}
which can be measured using time-of-flight or band-mapping techniques~\cite{bloch08}.

\begin{figure}[!t]
\includegraphics[width=0.99\columnwidth]{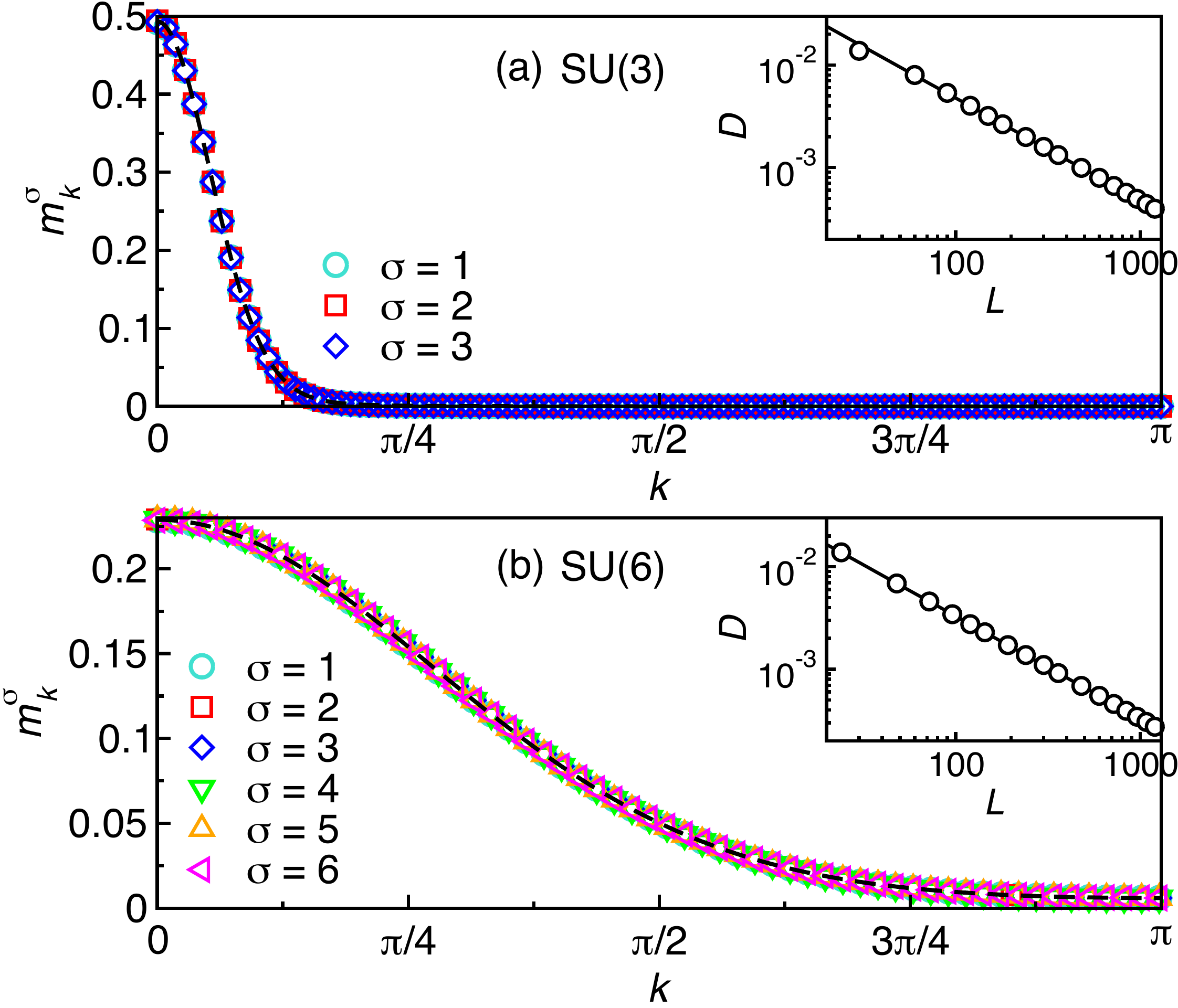}
\caption{Quasimomentum distribution function $m^\sigma_k$ in the ground state of: (a) the SU(3) model at filling $n=0.1$, and (b) the SU(6) model at filling $n=0.5$. Dashed lines are the averaged distributions $\overline{m}_k$. Calculations were done in chains with $L=1200$. Insets: Symbols show the corresponding average deviations $D$, defined in Eq.~(\ref{mom_diff}), plotted versus $L$. Solid lines are power-law ($\propto L^{-\alpha}$) fits to the data, where $\alpha=1.00$ for both.} 
\label{fig4}
\end{figure}

In Fig.~\ref{fig4}, we plot $m^\sigma_k$ for all flavors in open chains with 1200 sites. In Fig.~\ref{fig4}(a), we show results for the SU(3) case in systems at filling $n=N_p/L=0.1$, while in Fig.~\ref{fig4}(b) we show results for the SU(6) case at filling $n=0.5$. A remarkable property of $m^\sigma_k$ compared to $n^\sigma_l$ is that, in finite systems, the former is almost identical for all flavors despite the fact that the latter is not. The average carried by the sums in Eq.~\eqref{eq:nk} somehow erases the differences seen in $C^\sigma_l(x)$ for each $\sigma$ and $l$.

The dashed lines in Figs.~\ref{fig4}(a) and~\ref{fig4}(b) show the result for the average $m^\sigma_k$ over all spin flavors, $\overline{m}_k = \sum_\sigma m^\sigma_k/N = m_k/N$. As expected, the average follows the results for each value of $\sigma$. What is more interesting is to quantify how the differences between the curves for different flavors and the average change when one changes the system size. To do that, we compute the average deviation $D$
\begin{equation}
\label{mom_diff}
D = \sum_{\sigma=1}^{N} D^\sigma \, ,\ \ \text{where}\ \ D^\sigma = \frac{1}{2N_p} \sum_k |m^\sigma_k-\overline{m}_k| \,.
\end{equation}
The maximal possible value of $D$ is 1. 

In the insets in Figs.~\ref{fig4}(a) and~\ref{fig4}(b), we plot $D$ versus $L$ for chains with the same filling $n$ and number of flavors $N$ as in the main panels. These plots show that the average deviations are small already for small chains and decrease as a power law $L^{-1}$. The results in Fig.~\ref{fig4} make apparent that, even for small chains, one can accurately predict the quasimomentum distribution of each flavor using the total one-body correlations from Eq.~\eqref{avg_corr}.

\subsection{Extrapolations to the thermodynamic limit}\label{sec2_2}

Given that the power-law fits in the insets in Figs.~\ref{fig4}(a) and~\ref{fig4}(b) suggest that $m^\sigma_k$ becomes independent of $\sigma$ in the thermodynamic limit, here we study what happens to the one-body correlation matrix when one increases the system size. We focus on the case in which $N$ is $O(1)$, for which there is a well-defined filling per flavor $n^\sigma=N_p/(NL)$ in the thermodynamic limit. In our calculations we take $N \ll N_p$, for which robust finite-size scalings can be obtained.

\begin{figure}[!t]
\includegraphics[width=0.99\columnwidth]{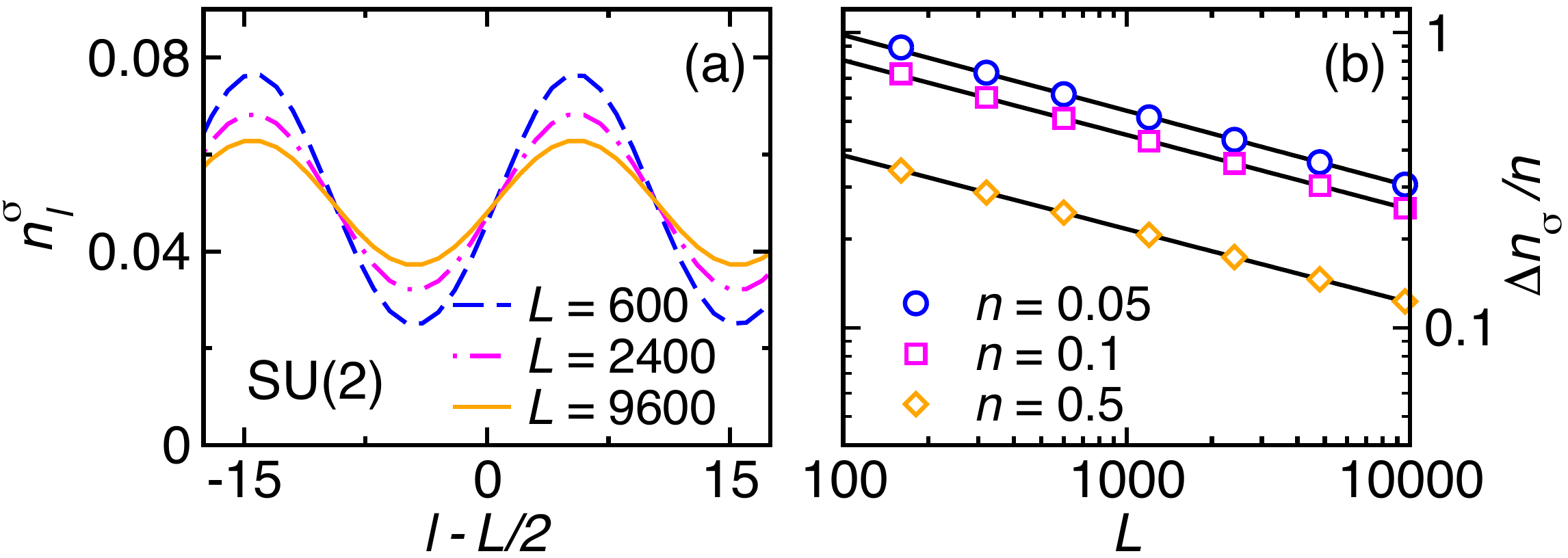}
\caption{(a) Ground-state site occupations $n^\sigma_l$ (for $\sigma=1$) in the SU(2) model at filling $n=0.1$, about the center of chains with different sizes. (b) Normalized difference $\Delta n_\sigma/n$ (for $\sigma=1$) between the site occupation at the peak and that at the dip closest to the chain center, for different fillings $n=0.05$, 0.1 and 0.5. Solid lines are power law fits $\Delta n_\sigma/n\propto L^{-\alpha}$ with $\alpha \simeq 0.25$.} 
\label{fig5}
\end{figure}

Let us first address what happens to the site occupations $n_l^\sigma$, shown in Fig.~\ref{fig2} for a finite chain, as one increases the chain size. For spinless fermions one knows that, in the thermodynamic limit away from the boundaries (after the Friedel oscillations have died out), the site occupations are position independent equal to $n=N_p/L$. 

\begin{figure}[!b]
\includegraphics[width=0.99\columnwidth]{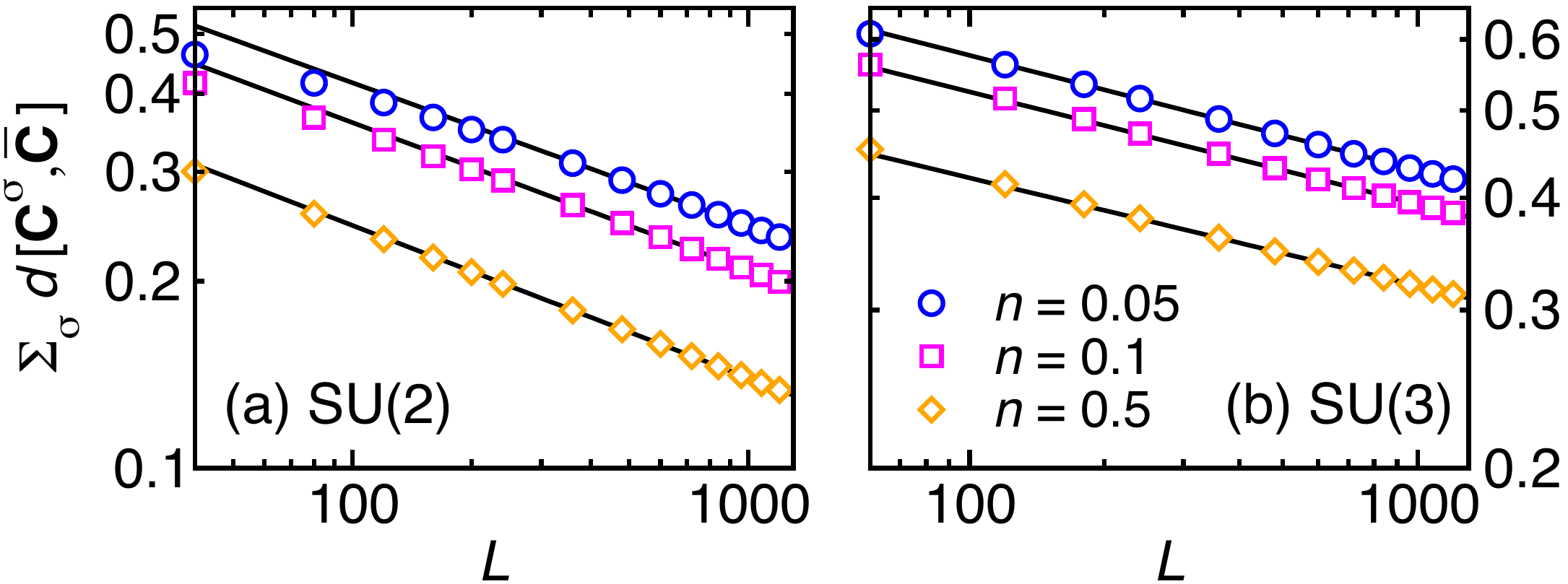}
\caption{Sums over trace distances $\sum_\sigma d[\mathbf{C}^\sigma,\overline{\mathbf{C}}]$ of the $\mathbf{C}^\sigma$ and $\overline{\mathbf{C}}$ one-body correlation matrices, where $d[\mathbf{C}^\sigma,\overline{\mathbf{C}}]$ is defined in Eq.~(\ref{tr_dist}). Results are shown for different fillings $n$ in the ground state of the (a) SU(2) and (b) SU(3) model. In all the cases the sums decrease as power laws. This is made apparent by the fits $\propto L^{-\alpha}$ depicted as solid lines, with $\alpha\simeq0.25$ (a) and $\alpha\simeq0.12$ (b).} 
\label{fig6}
\end{figure}

In Fig.~\ref{fig5}(a), we show $n_l^{\sigma=1}$ about the center of chains with three sizes, for the SU(2) model at a filling $n=0.1$. The differences between the maxima and the minima of $n_l^{\sigma=1}$ can be seen to decrease with increasing system size. To quantify them, we compute the difference between the site occupation at the peak $n^\sigma_{l_{\rm peak}}$ and that at the dip $n^\sigma_{l_{\rm dip}}$ that are closest to the lattice center, $\Delta n_\sigma=n^\sigma_{l_{\rm peak}}-n^\sigma_{l_{\rm dip}}$. Results for $\Delta n_{\sigma=1}/n$ for the SU(2) model at three values of $n$ are plotted in Fig.~\ref{fig5}(b) versus $L$. All three can be seen to decrease as power laws $\propto L^{-\alpha}$ with $\alpha\simeq0.25$. This suggests that, with increasing system size, the site occupations $n_l^{\sigma}$ become position independent away from the edges of the chain (as for spinless fermions) and are equal to $n^\sigma$. This means that the structures shown in $n^\sigma_l$ in Fig.~\ref{fig2}, which could be observed in experiments with ultracold fermions in optical lattices (in which $L\sim 100$), disappear in the thermodynamic limit away from the edges of the chain.

The results for $m^\sigma_k$ and $n_l^{\sigma}$ with increasing system size suggest that the one-body correlation matrices $\mathbf{C}^\sigma$ approach the average (over all flavors) one-body correlation matrix $\overline{\mathbf{C}} = \sum_{\sigma=1}^{N} \mathbf{C}^\sigma/N$ away from the edges of the chain. To verify this, we calculate the trace distances
\begin{equation}\label{tr_dist}
d[\mathbf{C}^\sigma,\overline{\mathbf{C}}]=\frac{1}{2N_p} \Tr \left\{\sqrt{[\mathbf{C}^\sigma-\overline{\mathbf{C}}]^2}\right\}\,,
\end{equation}
for all flavors. Figures~\ref{fig6}(a) and~\ref{fig6}(b) show $\sum_{\sigma=1}^{N} d[\mathbf{C}^\sigma,\overline{\mathbf{C}}]$ versus $L$ for the SU(2) and SU(3) models, respectively, at different fillings. With increasing system size, one can see that the added trace distances decrease as power laws in $L$. This suggests that the one-body correlation matrices $\mathbf{C}^\sigma$ are, up to nonextensive deviations (due to boundary effects), identical for all $\sigma$ in the thermodynamic limit.

However, it is important to stress that, in contrast to the results for $m^\sigma_k$ reported in Fig.~\ref{fig4}, the results for $n_l^{\sigma}$ in Fig.~\ref{fig5} and for $\sum_{\sigma=1}^{N} d[\mathbf{C}^\sigma,\overline{\mathbf{C}}]$ in Fig.~\ref{fig6} reveal that the spin-resolved one-body correlation functions can be quite different from the average in finite systems. These differences are likely not negligible for the system sizes relevant to ultracold-atom experiments. Moreover, the differences from the average increase with an increasing number of flavors $N$.

\subsection{Total one-body correlations} \label{sec2_3}

Since in the previous section we argued that the total one-body correlations $C_l(x)$ [divided by $N$, with $N$ being $O(1)$] become identical to the flavor-resolved ones $C^\sigma_l(x)$ in the thermodynamic limit, in what follows we focus our study on $C_l(x)$. The total one-body correlations $C_l(x)$ were introduced in Eqs.~(\ref{c_total}) and (\ref{avg_corr}) as an observable that highlights a universal property of the DQP model. In such a model, $C_l(x)$ depends neither on the number of flavors nor on the specific spin configuration (with contiguous fermions having distinct spin flavors). If one is interested in describing experiments, $C_l(x)$ may be good enough to describe quasimomentum distribution functions, but the calculation of $C^\sigma_l(x)$ may be needed to obtain accurate results for the spin-resolved site occupation profiles in small chains. 

\begin{figure}[!t]
\includegraphics[width=0.99\columnwidth]{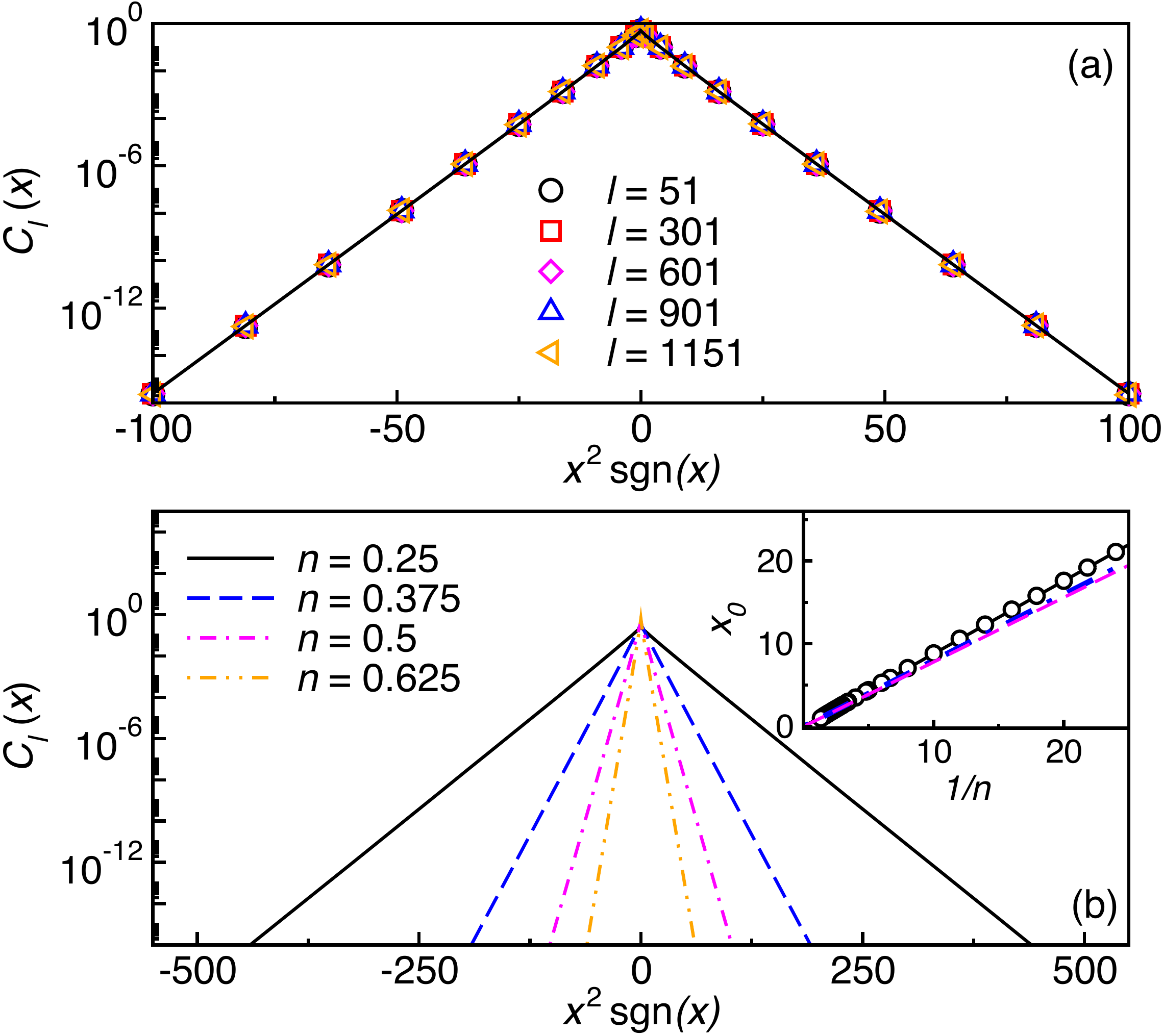}
\caption{Total one-body correlation function $C_l(x)$ in the ground state of a chain with $L=1200$. (a) $C_l(x)$ versus $x$ at filling $n=0.5$ for different values of $l$. There is data collapse for all values of $l$ shown. (b) $C_l(x)$ measured from the center of the lattice, $l=L/2+1$, for different fillings $n$. In all cases $C_l(x)$ can be seen to be Gaussian. Insets: Symbols depict the width $x_0$ of the Gaussian decay versus $1/n$. $x_0$ was obtained by fitting $C_l(x)$ to Eq.~(\ref{gaussian_peak}). The solid line is a linear fit to the data with slope 0.88, while the dashed and dotted lines represent $x_0^{\rm (SF)}$ from Eq.~(\ref{def_x0_sf}), and $x_0^{\rm (HO)}$ from Eq.~(\ref{def_x0_ho}), respectively.} 
\label{fig7}
\end{figure}

In Fig.~\ref{fig7}(a), we show the decay of $C_l(x)$ measured from different sites $l$, in a chain with $L=1200$ sites at filling $n=0.5$. The overlap between the results for different values of $l$ is nearly perfect, and the decay of $C_l(x)$ with $x$ is clearly Gaussian,
\begin{equation}
\label{gaussian_peak} C_l(x)=n \, e^{-x^2/x_0^2}\,,
\end{equation}
where $x_0$ is the width. Equation~(\ref{gaussian_peak}) is a defining property of the DQP model, and it is one of the main results of this work.

The Gaussian decay of the total one-body correlations $C_l(x)$ is a robust property of the DQP model. The robustness is characterized by three properties. First, as mentioned, Fig.~\ref{fig7}(a) shows that $C_l(x)$ measured at different sites $l$ yields nearly identical results even if $l$ is close to the boundaries of a finite chain. Second, Fig.~\ref{fig7}(b) shows that $C_l(x)$ is Gaussian for different chain fillings $n$. And third, Fig.~\ref{figapp1}, in Appendix~\ref{app1}, shows that $C_l(x)$ is independent of the system size $L$ for $L\gtrsim100$.

A key property of our setup, which we expect gives rise to the Gaussian decay of $C_l(x)$, is the distinguishability of the quantum particles [enforced by the projector $\hat{\cal M}_{l,x}$ defined in Eq.~(\ref{def_projM})]. Such a Gaussian decay is fundamentally different from the known power-law decay of one-body correlations of spinless fermions:
\begin{equation} \label{corr_sf}
\langle \hat c_{l+x}^\dagger \hat c^{}_l \rangle = \frac{\sin(n x \pi)}{x \pi} \, .
\end{equation}

It remains to be understood how the width $x_0$ of the Gaussian decay, Eq.~(\ref{gaussian_peak}), depends on the chain's filling. Dimensional analysis suggests that it is proportional to the average distance between particles, $x_0 \propto n^{-1}$. The proportionality constant can be estimated by assuming that the correlations in the DQP model and in the spinless fermion model approach each other when $n x\to 0$, i.e., at short distances when particle exchange ceases to play a role. Matching the second term in the expansion of Eqs.~(\ref{gaussian_peak}) and~(\ref{corr_sf}) about $x=0$ yields
\begin{equation} \label{def_x0_sf}
 x_0^{\rm (SF)} = \frac{1}{n} \frac{\sqrt{6}}{\pi} \,.
\end{equation}
We compare $x_0^{\rm (SF)}$ with the values of $x_0$ obtained by fitting $C_l(x)$ with the Gaussian function in Eq.~(\ref{gaussian_peak}). The results, shown in the inset in Fig.~\ref{fig7}(b), make apparent that $x_0$ is reasonably close to $x_0^{\rm (SF)}$.

Interestingly, the ground-state correlations of a single particle in a harmonic oscillator are also Gaussian. The ground-state wave function of such a system has the form $u_0(x) = \sqrt{n} e^{- (nx)^2\pi/2}$, where $n$ is the density at the center of the trap (related to the mass $m$ and the trap frequency $\omega$ by $n^2 \pi = m \omega/\hbar$). The correlation function $u_0(0) u_0(x)$ exhibits a Gaussian decay with a width 
\begin{equation} \label{def_x0_ho}
x_0^{\rm (HO)} = \frac{1}{n} \frac{\sqrt{2\pi}}{\pi}.
\end{equation}
$x_0^{\rm (HO)}$, which is also plotted in the inset in Fig.~\ref{fig7}(b), is very close to $x_0^{\rm (SF)}$ and is also close to $x_0$.

\begin{figure}[!t]
\includegraphics[width=0.99\columnwidth]{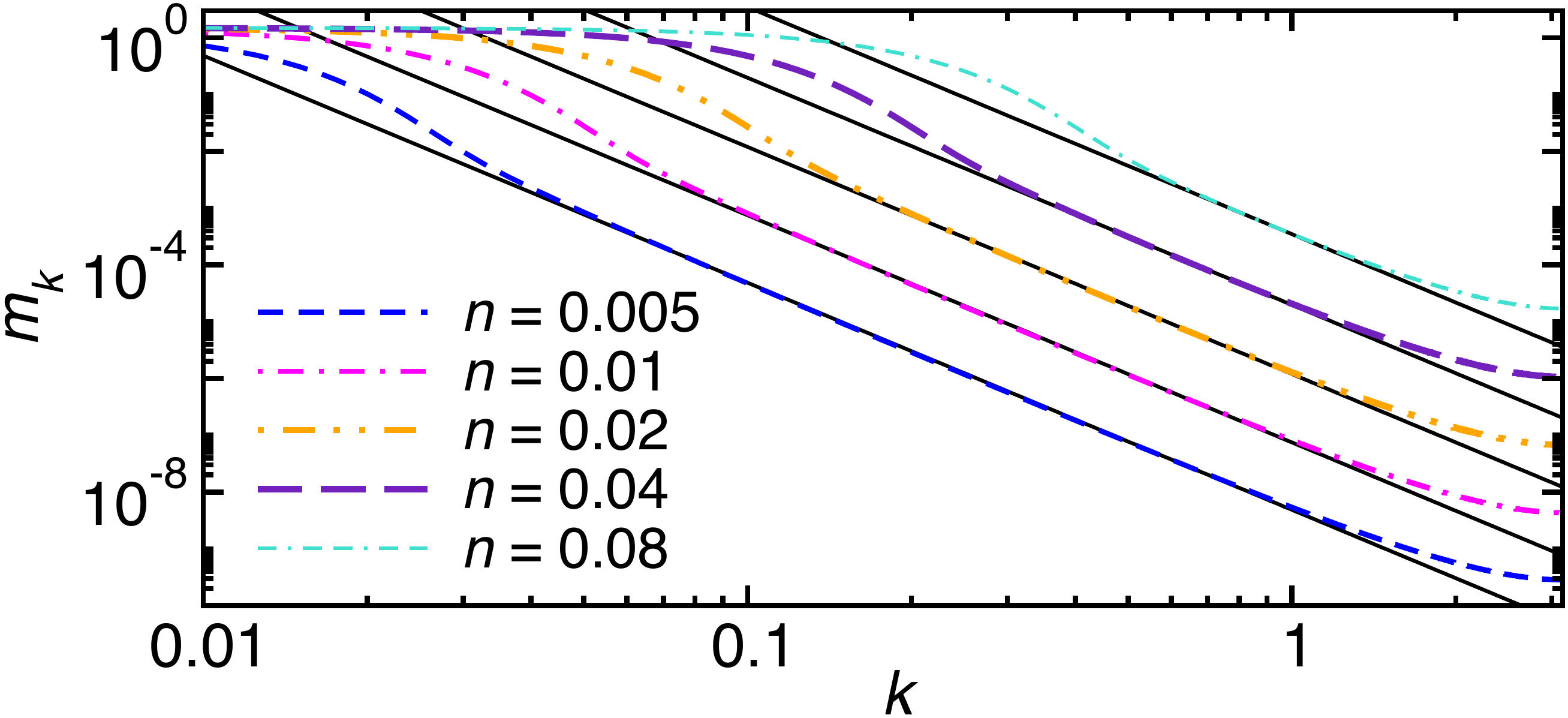}
\caption{High-momentum tails of the quasimomentum distribution $m_k$ of DQPs at low fillings in chains with $L=5000$ sites. Solid lines show fits to the expected $1/k^4$ behavior, which shrinks (and eventually disappears) with increasing filling as all momenta in the Brillouin zone become increasingly populated.} 
\label{fig8}
\end{figure}

Finally, related to the short-distance correlations of DQPs, it is important to note that, at low fillings, the quasimomentum distribution function $m_k$ of DQPs exhibits the $1/k^4$ tail that is known to appear in other models with contact interactions~\cite{cazalilla_citro_review_11}. In Fig.~\ref{fig8}, we plot $m_k$ vs $k$ in chains at low fillings. Fits to $1/k^4$ decay, depicted as solid lines, make apparent the region in $k$ in which the corresponding $1/k^4$ behavior occurs in $m_k$. Note that, with increasing filling, $m_k$ increases over the entire Brillouin zone and the region in which $1/k^4$ behavior occurs shrinks. It eventually disappears as all quasimomentum modes become significantly populated. Indications of such $1/k^4$ tails in impenetrable SU($N$) fermions in the continuum were recently reported in Ref.~\cite{jen_yip_18}. As for hard-core boson systems~\cite{rigol_muramatsu_04sept, xu_rigol_15}, our numerical approach in the lattice allows one to resolve those tails better than approaches that work directly in the continuum.

\section{Finite temperature} \label{sec4}

We now turn our attention to finite-temperature properties of the DQP model. We focus on the temperature dependence of the total one-body correlations $C_l(x; T)$. The total one-body correlations are shown in Sec.~\ref{sec3} to be universal, and to characterize flavor-resolved one-body correlations in the thermodynamic limit when $N$ is $O(1)$.

\subsection{Numerical implementation}

In order to compute the finite-temperature correlations of the DQP model, we develop a computational procedure similar to the one introduced for hard-core bosons in Ref.~\cite{rigol05_dec}.

The two basic relations needed to make the finite-temperature calculations in polynomial time are as follows~\cite{rigol05_dec}: 

(i) Traces over the fermionic Fock space of exponentials that are bilinear in fermionic creation and annihilation operators satisfy
\begin{eqnarray}
&&\mathrm{Tr} \left[ \exp\left( \sum_{ij}\hat c^\dagger_{i} X_{ij}\hat c^{}_{j}\right) 
\exp\left(\sum_{kl}\hat c^\dagger_{k} Y_{kl}\hat c^{}_{l} \right) \ldots \right] \nonumber\\
&& =\det\left[{\bf I}+ e^{\bf X}e^{\bf Y}\ldots\right].
\end{eqnarray}

(ii) The one-body operator $\hat c^\dagger_{l} \hat c^{}_{j}$, for $l\neq j$, can be written as
\begin{equation}\label{eq:Aij}
\hat c^\dagger_{l} \hat c^{}_{j}= \exp\left( \sum_{mn}\hat c^\dagger_{m} A_{mn}\hat c^{}_{n}\right) -1,
\end{equation}
where the only nonzero element in ${\bf A}$ is $A_{lj}=1$.

Using these two relations, one can write the total one-body correlation at finite temperature $C_l(x;T)$ as
\begin{equation}
\begin{aligned}
C_l(x\neq0;T)=&\frac{1}{Z}\{\det[\mathbf{I}+(\mathbf{I}+\mathbf{A})\mathbf{M}_{l,x}\mathbf{U}e^{-(\mathbf{E}-\mu\mathbf{I})/T}\mathbf{U}^\dagger] \\
&\quad -\det[\mathbf{I}+\mathbf{M}_{l,x}\mathbf{U}e^{-(\mathbf{E}-\mu\mathbf{I})/T}\mathbf{U}^\dagger]\} \, ,
\end{aligned}
\end{equation}
where $\bf{I}$ is the identity matrix, $\bf{A}$ is a matrix in which the only non-zero element is $A_{l,l+x}=1$ [from Eq.~\eqref{eq:Aij}], $\bf U$ is the unitary matrix that diagonalizes the corresponding spinless fermion Hamiltonian $\hat{H}_{\rm SF}$ in Eq.~(\ref{def_Hsf}), $\mathbf{H}_{\rm SF} = \mathbf{UEU}^\dagger$, with $\bf E$ being the diagonal matrix that contains all the single-particle eigenenergies, $Z=\prod_i[1+e^{-(E_{ii}-\mu)/T}]$ is the partition function, and $\mathbf{M}_{l,x}$ is the matrix representation of the projection operator $\hat{\cal{M}}_{l,x}$ [see Eq.~(\ref{def_projM})], which is a diagonal matrix with elements 0 between $l+1$ and $l+x-1$, and $1$ otherwise. 

The diagonal matrix elements of $C_l(x;T)$ are the same as for spinless fermions,
\begin{equation}
C_l(0;T)=1-[\mathbf{I}+e^{-(\mathbf{H}_{\rm SF}-\mu \mathbf{I})/T}]^{-1}_{ii}\,,
\end{equation}
and the chemical potential $\mu$ is determined so that the total number of particles $N_p=\sum_lC_l(0;T)$.

\subsection{Total one-body correlations}

\begin{figure}[!t]
\includegraphics[width=0.99\columnwidth]{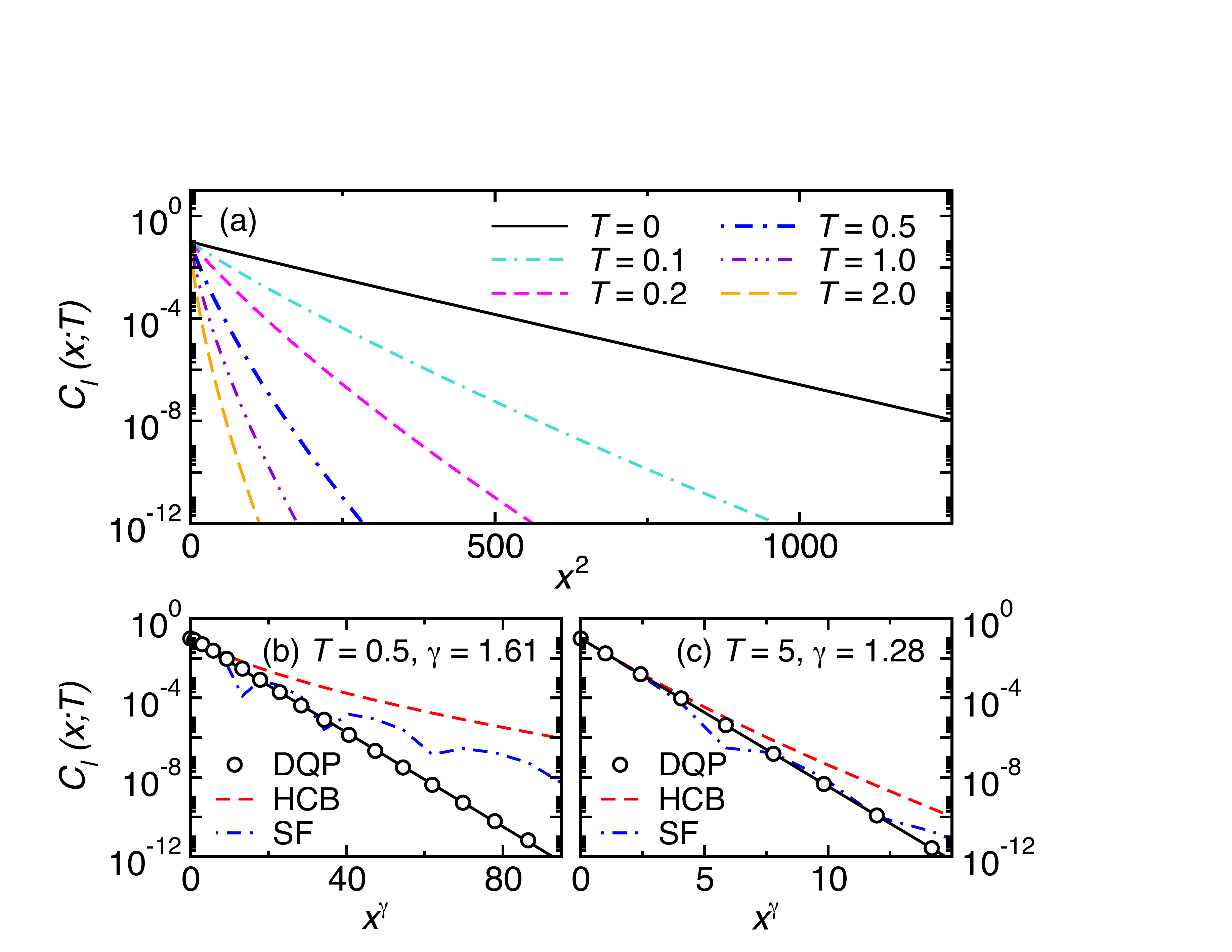}
\caption{Total one-body correlation function $C_l(x;T)$ at finite temperature in chains with $L=1200$, $l=L/2+1$, at filling $n=0.05$. (a) $C_l(x;T)$ in the DQP model as a function of $x^2$, for six temperatures. (b, c) Results for $T=0.5$ and 5, respectively, plotted as functions of $x^{\gamma(T)}$, where ${\gamma(T)}$ is the exponent extracted by fitting $C_l(x;T)$ with the ansatz in Eq.~(\ref{finiteT_corr}). Symbols show numerical results for the DQP model; solid lines, the corresponding fits. In (b) and (c), we compare the results obtained for the DQP model with those for the absolute value of the one-body correlations of spinless fermions (SF, dashed-dotted line) to which the DQP model is mapped. Results are also shown for hard-core bosons (HCB, dashed line), which can also be mapped onto the same SF Hamiltonian~\cite{cazalilla_citro_review_11}.} 
\label{fig9}
\end{figure}

In Fig.~\ref{fig9}(a) we plot the total one-body correlation function $C_l(x;T)$ versus $x^2$ for various temperatures. Figure~\ref{fig9}(a) shows that $\log_{10} C_l(x;T)$ becomes a convex function of $x^2$ at $T > 0$, which indicates that its decay is $\propto x^{\gamma(T)}$, with ${\gamma(T)}<2$. To describe the decay at finite temperatures, we use the fitting ansatz
\begin{equation}
\label{finiteT_corr}
C_l(x;T)=n\,\exp\{-[x/x_0(T)]^{\gamma(T)}\} \,,
\end{equation}
for which we determine the exponent $\gamma(T)$, and the effective width $x_0(T)$, as functions of the temperature. We fit $\log_{10} C_l(x;T)$ from $x=0$ through all the sites in which $C_l(x;T)\geq 10^{-12}$, and choose temperatures such that the fitting includes at least six points. The latter constrains the highest temperatures for which we do fits.

Examples of fits using Eq.~(\ref{finiteT_corr}) are reported in Figs.~\ref{fig9}(b) and~\ref{fig9}(c) for temperatures $T=0.5$ and 5, respectively, in systems with $n=0.1$. [The numerical results for $C_l(x;T)$ are shown as symbols and the fits are shown as solid lines.] Note the near-perfect overlap between the numerical results and the fits, as well as the fact that $\log_{10} C_l(x;T)$ versus $x^{\gamma(T)}$ exhibits a linear decrease when the appropriate value of ${\gamma(T)}$ is used; i.e., these plots make apparent that the ansatz in Eq.~(\ref{finiteT_corr}) provides an accurate description of the total one-body correlations at finite temperature.

The values of ${\gamma(T)}$ obtained in Figs.~\ref{fig9}(b) and~\ref{fig9}(c) suggest that $\log_{10} C_l(x;T)$ approaches a linear function of $x$ as $T$ increases. This is consistent with the intuition that, at very high temperature, the statistics of the particles ceases to play a role and one-body correlations of impenetrable SU($N$) fermions should become identical to those of the spinless fermions to which they are mapped, for which an exponential decay is known to occur at finite temperature. In Figs.~\ref{fig9}(b) and~\ref{fig9}(c), we also show the one-body correlations for the corresponding spinless fermions and hard-core bosons (see Appendix~\ref{app3} for the definition of hard-core bosons). Hard-core bosons are also mappable to the spinless fermion Hamiltonian to which we mapped the constrained impenetrable SU($N$) fermions~\cite{cazalilla_citro_review_11}. As expected, the results for all three models approach each other with increasing temperature. 

\begin{figure}[!t]
\includegraphics[width=0.99\columnwidth]{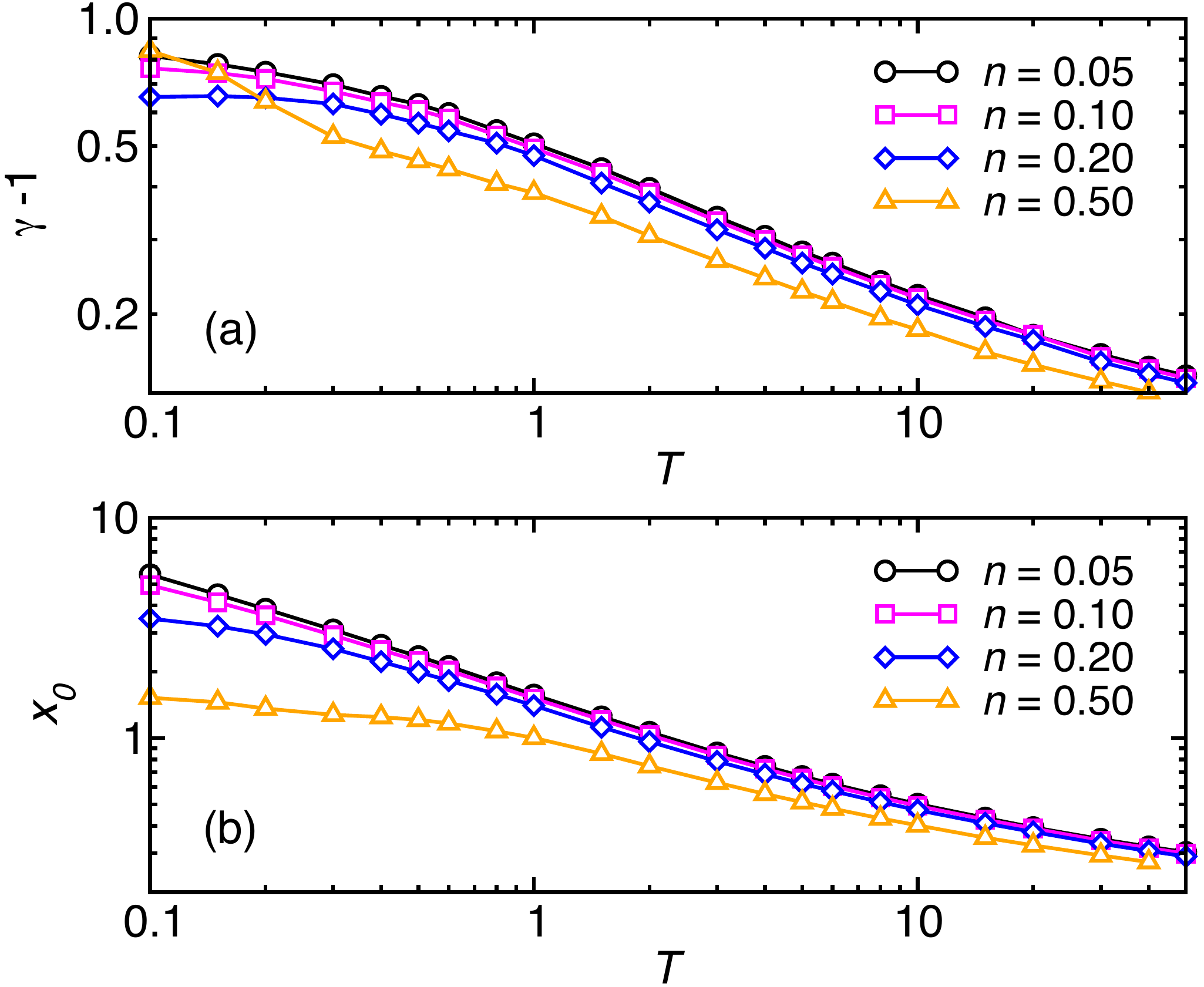}
\caption{(a) ${\gamma(T)}$ and (b) $x_0(T)$, obtained by fitting our numerical results with Eq.~(\ref{finiteT_corr}), plotted as functions of $T$ for different fillings $n$. Symbols show the numerical results, while solid lines are guides for the eye.} 
\label{fig10}
\end{figure}

Figure~\ref{fig10}(a) shows how $\gamma(T)$ approaches $1$ with increasing temperature for different fillings $n$. It is interesting to note that, despite the fact that the exponent of the stretched exponential decreases with increasing temperature, Fig.~\ref{fig9}(a) shows that the higher the temperature the smaller the correlations at any given distance $x$, for $C_l(x;T)\geq 10^{-12}$. This occurs because, as shown in Fig.~\ref{fig10}(b), $x_0(T)$ also decreases with increasing temperature. We should add that the departure of $C_l(x;T)$ from a Gaussian at finite temperature results in an enhancement of one-body correlations at long distances with respect to the ground state (see Appendix~\ref{app2}). This is something that may be of experimental interest at low temperatures. 

\begin{figure}[!t]
\includegraphics[width=0.99\columnwidth]{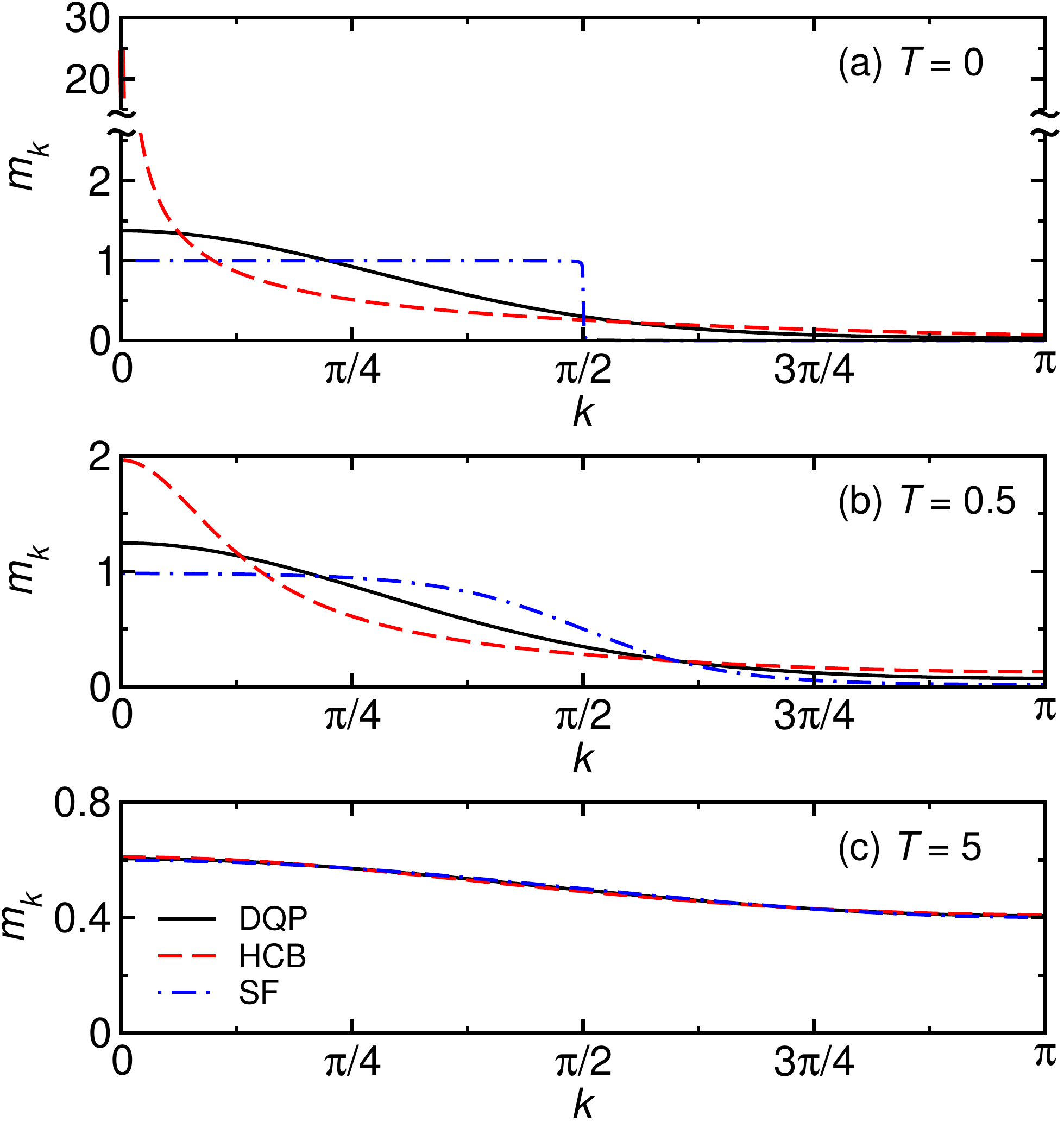}
\caption{Quasimomentum distribution function $m_k$ of distinguishable quantum particles (DQP, solid lines), spinless fermions (SF, dashed-dotted lines), and hard-core bosons (HCB, dashed lines) at (a) $T=0$, (b) $T=0.5$, and (c) $T=5$. Results were obtained in chains with $L=1200$ sites at filling $n=0.5$.} 
\label{fig11}
\end{figure}

To conclude, we report results for the total quasimomentum distribution function $m_k(T) = \sum_{l,x}e^{-ikx}C_l(x;T)/L$, which is of special interest for experiments with ultracold fermions. $m_k(T)$ for the DQP model is shown in Fig.~\ref{fig11} at three temperatures, $T=0$, 0.5, and 5. In this figure, we also show the quasimomentum distribution functions of spinless fermions and hard-core bosons at the same temperatures.

In the ground state, $m_k$ of DQPs shows a smooth peak near $k=0$ (see also Fig.~\ref{fig4}). This is in stark contrast to the quasimomentum distribution of spinless fermions, which exhibits a step like distribution with a Fermi edge, and of hard-core bosons, which exhibits a sharp peak at $k=0$ (note the discontinuity in the $y$ axis), making apparent the occurrence of quasicondensation~\cite{cazalilla_citro_review_11, rigol_muramatsu_04sept, rigol_muramatsu_05july}. Temperatures below the energy scale of the hopping [see Fig.~\ref{fig11}(b)] do not change the quasimomentum distribution of DQPs much, change the quasimomentum distribution of spinless fermions about the Fermi edge, and have a dramatic effect on the quasimomentum distribution of hard-core bosons. The latter occurs because one-body correlations switch from power-law to exponential decay when $T$ becomes nonzero~\cite{rigol05_dec}. At temperatures above the bandwidth of the spinless fermion model (4 in our units) [see Fig.~\ref{fig11}(c)], the quasimomentum distributions of DQPs, spinless fermions, and hard-core bosons become near indistinguishable. This, which is consistent with the results for $C_l(x;T)$ shown in Fig.~\ref{fig9}(c), highlights the irrelevance of the particle statistics in $m_k$ at these temperatures.

\section{Summary} \label{sec5}

We have studied impenetrable SU($N$) fermions within sectors of the Hamiltonian in which consecutive fermions have different spin flavors. We call this constrained model a model of distinguishable quantum particles (DQPs), for which the original statistics of the particles plays no role. This is because contiguous particles have different spin flavors and particle exchanges are forbidden by the impenetrability constraint. Consequently, our results apply equally to impenetrable SU($N$) bosons under the same constraint that contiguous bosons have different spin flavors. For the model of DQPs, we have introduced an exact numerical approach based on a mapping onto noninteracting spinless fermions that allows one to compute spin-resolved one-body correlation functions in eigenstates of the Hamiltonian and at finite temperature.

We have shown that, in the ground state of the DQP model, the decay of one-body correlations is Gaussian. This is in contrast to the power-law or exponential decay known to occur in the ground state of traditional 1D models~\cite{cazalilla_citro_review_11}. We have also shown that, at low fillings in the lattice, the quasimomentum distribution function of DQPs exhibits a $1/k^4$ tail. At finite temperatures, we have shown that one-body correlations are well described by a stretched exponential decay, with an exponent that transitions between 2 and 1 as the temperature increases. Namely, the correlations transition between Gaussian in the ground state and exponential at high temperatures. At high temperatures, we have also shown that the momentum distribution function of DQPs becomes identical to those of spinless fermions and hard-core bosons. 

As an outlook, it would be interesting to find other 1D models in which one-body correlations exhibit Gaussian decay in the ground state. This might help shed further light on the conditions needed for such correlations to occur and on the universality of our results.

\section{Acknowledgments}
We acknowledge discussions with T. \v Cade\v z, M. Endres, P. Prelov\v sek, and T. Prosen. Y.Z. and M.R. acknowledge support from NSF Grant No.~PHY-1707482. Y.Z. acknowledges support from an Advanced Grant of the European Research Council (ERC; No. 694544-OMNES). L.V. acknowledges support from the Slovenian Research Agency (Research Core Funding No. P1-0044).

\appendix

\section{Finite size effects of $C_l(x)$ in the ground state} \label{app1}

\begin{figure}[!t]
\includegraphics[width=0.99\columnwidth]{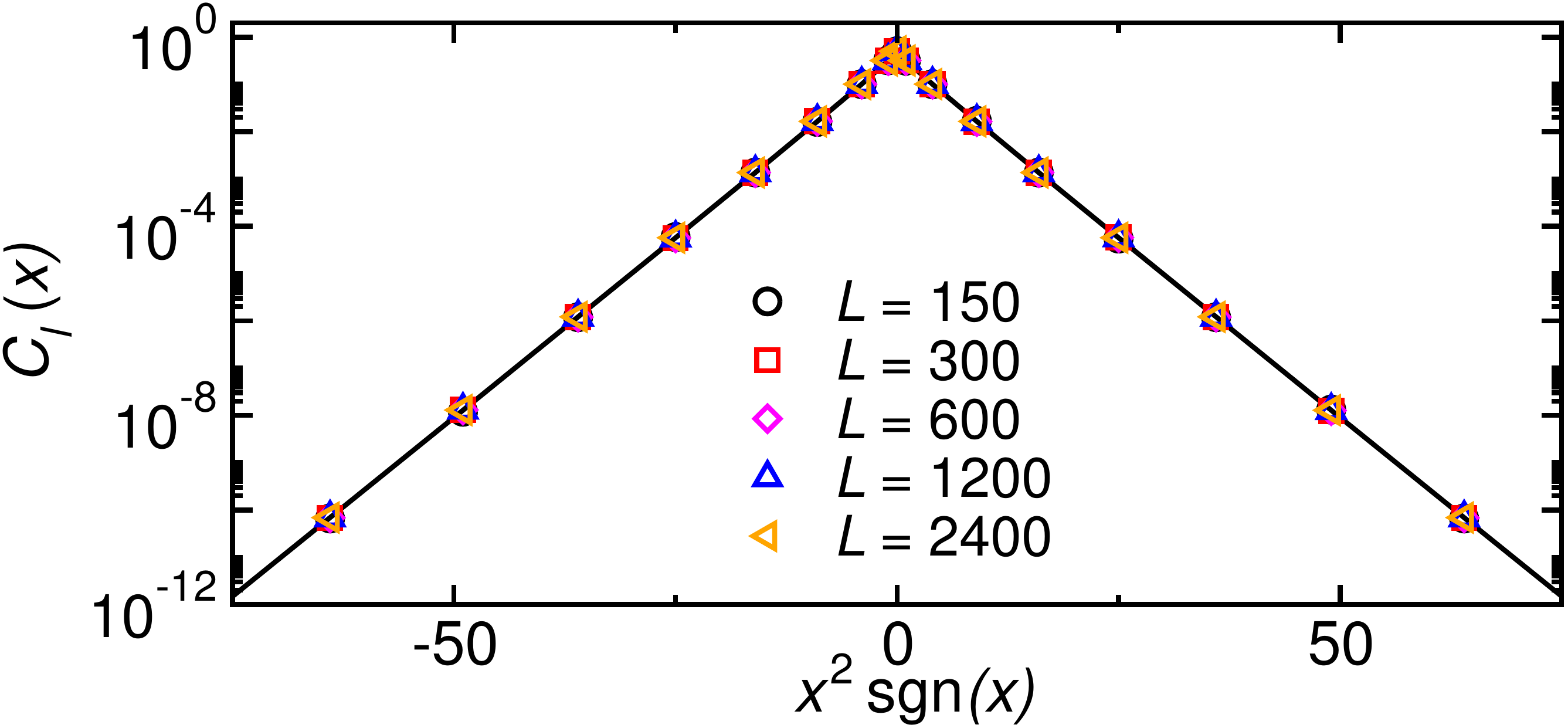}
\caption{Total one-body correlation function $C_l(x)$ in the ground state of five chains of different sizes. The filling is fixed to $n=0.5$, and $l$ is chosen to be $l=L/2+1$. The results for different chain sizes $L$ are virtually indistinguishable.} 
\label{figapp1}
\end{figure}

In Fig.~\ref{figapp1}, we show $C_l(x)$ for five chains of different sizes $L$ at filling $n=0.5$. The results for $C_l(x)$ agree with each other independently of the values of $L$ chosen, which means that finite-size effects for the total one-body correlations are negligible already for systems with $L\sim 100$. This is in stark contrast to the spin-resolved correlations $C^\sigma_l(x)$ (see Fig.~\ref{fig6}), which can exhibit significant finite-size effects for much larger chain sizes. 

In Fig.~\ref{fig7}, we chose $L=1200$ for the calculations of $C_l(x)$. For this chain size, we expect the numerical results to be indistinguishable from those in the thermodynamic limit.

\section{Hard-core bosons} \label{app3}

The hard-core boson Hamiltonian can be written as
\begin{equation} \label{def_Hhcb}
 \hat H_{\text{HCB}} = -  \sum_{l=1}^{L-1} \left(\hat b_{l+1}^\dagger \hat b^{}_l + {\rm H.c.} \right) \, ,
\end{equation}
supplemented by the constraints $(\hat b^{}_{l})^2 = (\hat b_{l}^\dagger)^2 = 0$, where $\hat b_{l}^\dagger$ ($\hat b^{}_l$) is the creation (annihilation) operator of a hard-core boson at site $l$. This model is the infinite on-site repulsion limit of the Bose-Hubbard model~\cite{cazalilla_citro_review_11}. By virtue of the Holstein-Primakoff and the Jordan-Wigner transformations~\cite{jordan_wigner_28, holstein_primakoff_40, cazalilla_citro_review_11}, one can map the Hamiltonian in Eq.~(\ref{def_Hhcb}) onto a Hamiltonian of spinless fermions, Eq.~(\ref{def_Hsf}), using $\hat b_l = e^{i \pi \sum_{m<l} \hat c_m^\dagger \hat c_m} \hat c_l $. We calculate the one-body correlations of hard-core bosons using the approach introduced in Refs.~\cite{rigol_muramatsu_04sept} and \cite{rigol_muramatsu_05july} for the ground state, and in Ref.~\cite{rigol05_dec} for finite temperatures.

\section{Low temperature behavior of $C_l(x;T)$} \label{app2}

\begin{figure}[!t]
\includegraphics[width=0.99\columnwidth]{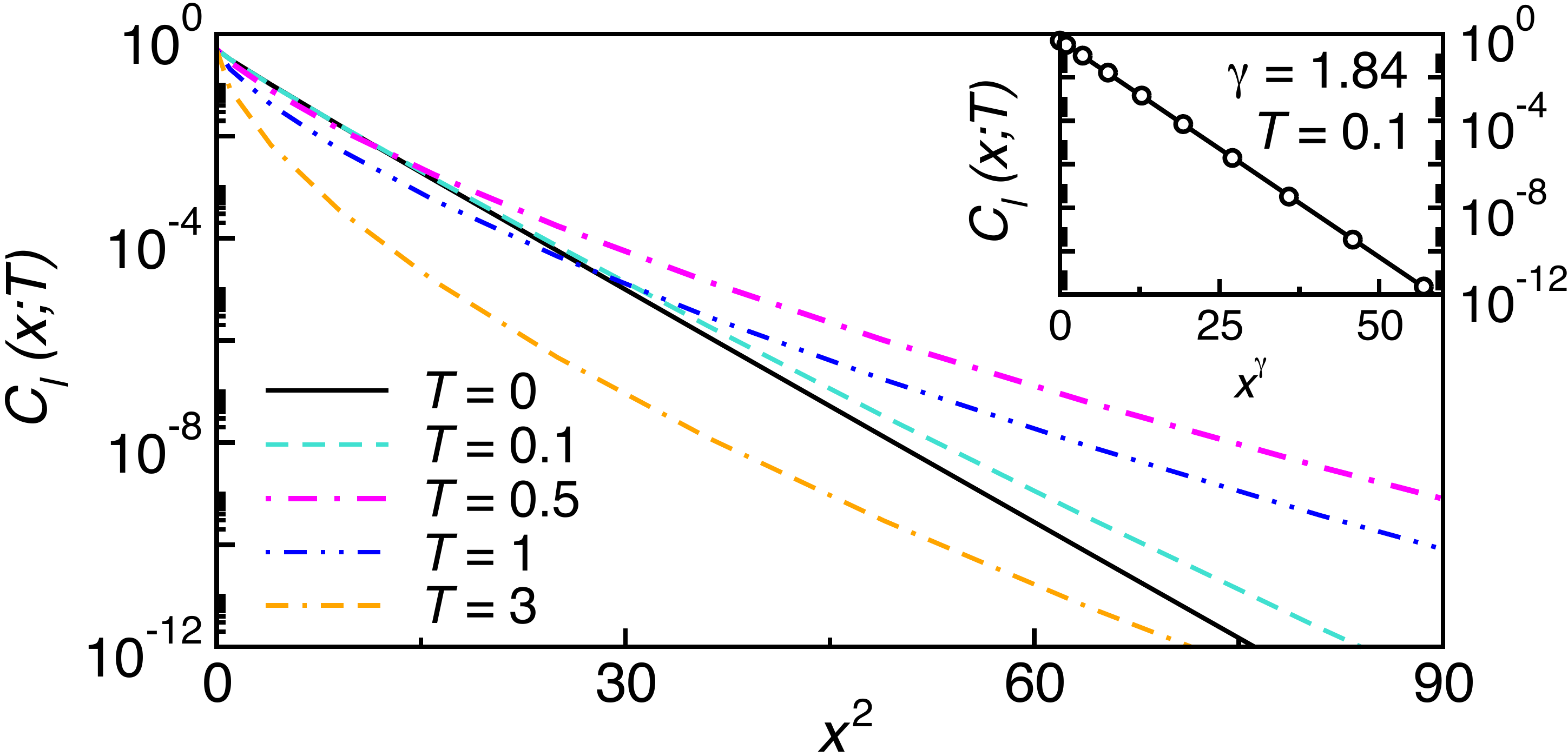}
\caption{Total one-body correlations $C_l(x;T)$ at low and intermediate temperatures for $n=0.5$, $L=1200$, and $l=L/2+1$. Inset: Symbols depict results for $T=0.1$ (also shown in the main panel), while the solid line is a fit to Eq.~(\ref{finiteT_corr}) with $\gamma(T)=1.84$.} 
\label{figapp2}
\end{figure}

Figure~\ref{figapp2} shows the low- and intermediate-temperature behavior of $C_l(x;T)$ versus $x$ at filling $n=0.5$. The main point to be highlighted about these results is that while finite temperatures always reduce the total one-body correlations at short distances, the switch from Gaussian in the ground state to stretched exponential decay at finite temperature (see the inset) results in an enhancement of the total one-body correlations at long distances. This enhancement is likely to be relevant to experiments only at low temperatures, so that the correlations are not too small to be detected. Such finite-temperature behavior at long distances, not apparent in the occupations of the low-$k$ momenta, which decrease with increasing temperature [see Figs.~\ref{fig11}(a) and~\ref{fig11}(b)], is another remarkable property of DQPs compared to traditional one-dimensional models. In the latter models, one-body correlations at long distances are usually reduced at finite temperatures with respect to the ground state.

\newpage

\bibliographystyle{biblev1}
\bibliography{references}

\end{document}